\documentclass[useAMS,usenatbib]{mn2e}
\usepackage[dvips]{graphicx}
\usepackage{lscape}

\title[The Ubiquity of AGN Activity in the local Universe]
{Towards a Complete Census of AGNs in Nearby Galaxies: \\
A Large Population of Optically Unidentified AGNs}
\author[A.~D. Goulding \& D.~M. Alexander]
  {A.D.~Goulding \thanks{E-mail:- andrew.goulding@durham.ac.uk} \&
    D.M.~Alexander \\
  Department of Physics, Durham University, South Road, Durham.}
\date{Released 2009 Xxxxx XX}

\pagerange{\pageref{firstpage}--\pageref{lastpage}} \pubyear{2008}

\def\LaTeX{L\kern-.36em\raise.3ex\hbox{a}\kern-.15em
    T\kern-.1667em\lower.7ex\hbox{E}\kern-.125emX}

\def\chandra{{\it Chandra}}
\def\iras{{\it IRAS}}

\def\spitzer{{\it Spitzer}}

\def\cm{{\rm\thinspace cm}}

\def\erg{{\rm\thinspace erg}}

\def\km{{\rm\thinspace km}}

\def\Lsun{\hbox{$\rm\thinspace L_{\odot}$}}

\def\Msun{\hbox{$\rm\thinspace M_{\odot}$}}

\def\s{{\rm\thinspace s}}

\def\ergpcmsqps{\hbox{$\erg\cm^{-2}\s^{-1}\,$}}

\def\ergps{\hbox{$\erg\s^{-1}\,$}}

\def\kmps{\hbox{$\km\s^{-1}\,$}}

\def\pcm{\hbox{$\cm^{-3}\,$}}
\def\pcmsq{\hbox{$\cm^{-2}\,$}}

\def\Lir{\hbox{${\rm L}_{\rm IR}$}}
\def\Lnev{\hbox{${\rm L}_{\rm [NeV]}$}}

\def\um{\hbox{$\ \umu {\rm m}$}}
\def\nev{[NeV] $\lambda 14.32 \um$ }
\def\oiv{[OIV] $\lambda 25.89 \um$ }

\begin{document}

\label{firstpage}

\maketitle

\begin{abstract}
Using {\it Spitzer}-IRS spectroscopy, we investigate the ubiquity of
Active Galactic Nuclei (AGN) in a complete ($\approx 94$ percent),
volume-limited sample of the most bolometrically-luminous galaxies
(${\rm L}_{\rm IR, \ 8-1000 \um} \ga$~(0.3--20)~$\times 10^{10}
\Lsun$) to $D < 15$~Mpc. Our analyses are based on the detection of
the high-excitation emission line [NeV]$\lambda 14.32 \um$ (97.1 eV)
to unambiguously identify AGN activity. We find that 17 of the 64
IR-bright galaxies in our sample host AGN activity ($\approx
27^{+8}_{-6}$ percent), \hbox{$\goa 50$} percent of which are not
identified as AGNs using optical spectroscopy. The large AGN fraction
indicates a tighter connection between AGN activity and IR luminosity
for galaxies in the local Universe than previously found, potentially
indicating a close association between AGN activity and star
formation. The optically unidentified AGNs span a wide range of galaxy
type (S0--Ir) and are typically starburst-dominated systems hosting
modest-luminosity AGN activity ($\Lnev \approx 10^{37}$--$10^{39}
\ergps$). The non-identification of optical AGN signatures in the
majority of these galaxies appears to be due to extinction towards the
AGN, rather than intrinsically low-luminosity AGN activity; however,
we note that the AGN optical signatures are diluted in some galaxies
due to strong star-formation activity. Examination of optical images
shows that the optically unidentified AGNs with evidence for
extinction are hosted in either highly inclined galaxies or galaxies
with dust lanes, indicating that obscuration of the AGN is not
necessarily due to an obscuring torus. We therefore conclude that
optical spectroscopic surveys miss approximately half of the AGN
population simply due to extinction through the host galaxy.


 \end{abstract}


\section{Introduction}

The seminal discovery that all massive galaxies in the local Universe
harbour super-massive black holes (SMBHs; $M_{\rm
BH}>10^6$~$M_{\odot}$) implies that all massive galaxies have hosted
Active Galactic Nuclei (AGN) at some time during the last
$\approx$~13~Gyrs (e.g., \citealt{kormendy95}; \citealt{magorrian98};
\citealt{gebhardt00}). Sensitive blank-field surveys have traced the
evolution of luminous AGN activity out to $z\approx$~5--6, providing a
window on the growth of SMBHs across $\approx$~95\% of cosmic time
(e.g.,\ \citealt{ueda03}; \citealt{croom04}; \citealt{fan04};
\citealt{hasinger05}; \citealt{richards06}). However, these
blank-field surveys typically lack the volume and sensitivity to
provide a complete census of AGN activity in the local Universe. Such
a census is required to (1) provide a baseline with which to interpret
the results obtained for distant AGNs from blank-field surveys, and
(2) provide definitive constraints on the growth of SMBHs in the local
Universe (e.g.,\ the growth rates of SMBHs; the relative amount of
obscured and unobscured SMBH growth; the galaxies and environments
where SMBHs are growing).

Arguably, the most complete census of AGN activity in the local
Universe is the optical spectroscopic survey of
\citeauthor{ho97a}(\citeyear{ho97b}a,b; hereafter Ho97). Ho97
classified nearly all galaxies with $B_T\le12.5$ mag in the Northern
hemisphere and identified AGNs on the basis of the relative strength
of forbidden and permitted emission lines (e.g.,\ \citealt{bpt};
\citealt{veil_oster87}; \citealt{kewley_bpt}); objects with a broad
permitted line component (FWHM $> 1000\kmps$) were also classified as
AGNs. Using this classification scheme, Ho97 found that $\approx$~10\%
of galaxies unambiguously host AGN activity (i.e.,\ optically
classified as Seyfert galaxies). These AGNs were found to
predominantly reside in moderately massive bulge-dominated galaxies
(Hubble type E--Sbc). However, since the source selection and
classification were performed at optical wavelengths, the Ho97 studies
were insensitive to the identification of the most heavily
dust-obscured AGNs. For example, the nearby Scd galaxy, NGC~4945 is
classified as a starburst galaxy at optical wavelengths and only
reveals the presence of AGN activity using mid-IR spectroscopy
\citep{spoon2000} and X-ray observations \citep{n4945}. Various
studies have suggested that NGC~4945 is unlikely to be a particularly
unusual AGN (e.g.,\ \citealt{lutz03}; \citealt{maiolino03}),
indicating that there may be many more optically unidentified AGNs in
the local Universe.

X-ray observations have revealed potential AGNs in many galaxies in
the Ho97 sample lacking optical AGN signatures (e.g.,\ \citealt{ho01};
\citealt{desroches09}). However, there is often ambiguity over whether
an AGN is producing the X-ray emission in these galaxies (e.g., there
can be significant contamination from X-ray binaries). Furthermore,
the X-ray emission from Compton-thick AGNs ($N_{\rm
H}>1.5\times10^{24}$~cm$^{-2}$) can be extremely weak, making it
challenging to identify the most heavily obscured AGNs using X-ray
data alone (i.e.,\ the $<10$~keV emission can be a factor
$\approx$~30--1000 times weaker than the intrinsic emission; e.g.,\
\citealt{risaliti99}; \citealt{matt2000}). By comparison, due to the
extreme conditions required to produce the high ionization emission
line [NeV] $\lambda 14.32$, $24.32\um$ (97.1~eV), mid-IR spectroscopy
provides an unambiguous indicator of AGN activity in nearby galaxies
(e.g.,\ \citealt{weedman05}; \citealt{armus06}).\footnote{We note that
[OIV]~$\lambda$~25.9~$\mu$m (54.9~eV) is also often used for AGN
identification, although energetic starbursts can also produce
luminous [OIV] emission; see \S3.4.} The relative optical depth at
mid-IR wavelengths is also considerably lower than at optical
wavelengths ($A_{\rm \lambda 14.3 \um}$/$A_{\rm V}$~$\approx$~50;
\citealt{li01}). Even heavily Compton-thick AGNs that are weak at
X-ray energies can be identified using mid-IR spectroscopy (e.g.,\ the
well-studied NGC~1068 with $N_{\rm H} > 10^{25} \pcmsq$ has bright
\nev emission; \citealt{sturm02}). Thus, the identification of
high-ionization [NeV] $\lambda 14.32$, $24.32\um$ provides a
relatively optically thin means by which to probe the central engine
of nearby AGNs. On the basis of the identification of [NeV] emission
in a heterogeneous sample of late-type spiral galaxies within the
local Universe, \citeauthor{sat08} (2008; hereafter, S08) have
suggested that a large number of Sc--Sm galaxies host optically
unidentified AGN activity.

Local mid-IR surveys (e.g., \citealt{sturm02}; the \spitzer \ Infrared
Nearby Galaxy Survey [SINGS] Legacy Project [\citeauthor{dale06} 2006;
hereafter D06] and S08) have shown the advantages of using mid-IR
spectroscopy as an AGN diagnostic. However, none of these studies have
used this diagnostic to provide a complete unambiguous census of AGN
activity within the local Universe. Here we use sensitive
high-resolution \spitzer-IRS spectroscopy to identify AGNs within a
complete ($\approx 94$ percent) volume-limited survey of the most
bolometrically luminous galaxies ($\Lir > 3 \times 10^9 \Lsun$) to a
distance of $D < 15$ Mpc.\footnote{$\Lir$ corresponds to the
8--1000~$\mu$m luminosity, as defined by \citet{sanders96}.} By
selecting galaxies at IR wavelengths, our sample will comprise the
most active galaxies in the local Universe and will also include the
most dust-obscured systems. We use these data to unambiguously
identify AGNs using the high-ionization \nev emission line to produce
the most sensitive census of AGN activity in the local Universe to
date. In \S2 we outline the construction and data reduction analysis
of the sample assembled from the {\it IRAS} Revised Bright Galaxy
Survey of Sanders et~al. (2003; RBGS). In \S3 we determine the
fraction of local galaxies hosting AGN activity, explore their
properties, and compare the results to the previous optical survey of
Ho97 to address the key question: Why are a large number of AGNs
unidentified at optical wavelengths? In \S4 we present our
conclusions.

\begin{figure}
\hspace{-0.5cm}
\includegraphics[width=1.1\linewidth]{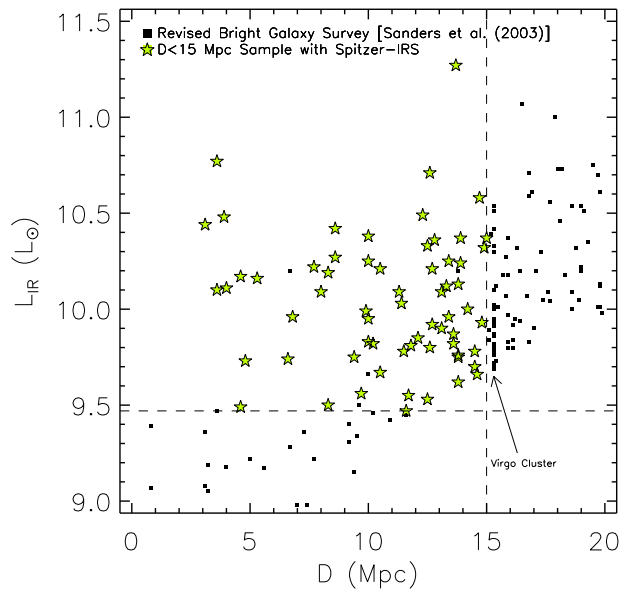}
\caption{Logarithm of IR luminosity versus luminosity distance for all
objects in the RBGS (\citealt{RBGS}; squares). The 64 IR-bright
galaxies ($\Lir \approx 3 \times 10^9 \Lsun$) to $D < 15$ Mpc with
high-resolution \spitzer-IRS spectroscopy are explored here
(stars).}
\label{fig_1}
\end{figure}

\section{The Sample and Data Reduction}
\subsection{Sample Selection}
Using \iras, the RBGS (\citealt{RBGS}) has provided an accurate census
of all IR-bright galaxies ($|b| > 5 \degr$, $f_{60 \mu m} > 5.24$ Jy)
in the local Universe. The aim of our study is to identify AGN
activity in the most bolometrically luminous galaxies ($\Lir > 3
\times 10^9 \Lsun$) out to $D < 15$ Mpc.\footnote{Distances have been
calculated using the cosmic attractor model of \citet{mould00}, which
adjusts heliocentric redshifts to the centroid of the local group,
taking into account the gravitational attraction towards the Virgo
cluster, the Great Attractor, and the Shapley supercluster.} The
distance constraint of 15~Mpc was placed so as to not include the
Virgo cluster at 16~Mpc (i.e.,\ to be representative of field-galaxy
populations). The IR luminosity threshold was chosen to be well
matched to the flux limit of the RBGS (see Fig.~\ref{fig_1}) and
ensures that we do not include low-luminosity dwarf galaxies and
relatively inactive galaxies. In the RBGS there are 68 \iras \
detected galaxies to a

\begin{landscape}

\begin{table}
\begin{minipage}{240mm}
\begin{center}
\vspace{1.0cm}
\caption{Catalogue of sources and derived quantities.}
\begin{tabular}{lccrcccrccccccc}
\hline
  \multicolumn{1}{c}{Common} &
  \multicolumn{1}{c}{RA} &
  \multicolumn{1}{c}{DEC} &
  \multicolumn{1}{c}{Dist} &
  \multicolumn{1}{c}{SH} &
  \multicolumn{1}{c}{LH} &
  \multicolumn{1}{c}{Morph} &
  \multicolumn{1}{c}{$\Lir$} &
  \multicolumn{1}{c}{Spectral} &
  \multicolumn{1}{c}{\underline{$H_{\alpha}$}} &
  \multicolumn{1}{c}{\underline{[OIII]}} &
  \multicolumn{1}{c}{\underline{[SII]}} &
  \multicolumn{1}{c}{\underline{[NII]}} &
  \multicolumn{1}{c}{$10^{15} f_{\rm{[OIII]}}$} &
  \multicolumn{1}{c}{Ref.} \\
  \multicolumn{1}{c}{Name} &
  \multicolumn{1}{c}{} &
  \multicolumn{1}{c}{} &
  \multicolumn{1}{c}{(Mpc)} &
  \multicolumn{1}{c}{(kpc)} &
  \multicolumn{1}{c}{(kpc)} &
  \multicolumn{1}{c}{Class} &
  \multicolumn{1}{c}{($L_{\odot}$)} &
  \multicolumn{1}{c}{Class} &
  \multicolumn{1}{c}{$H_{\beta}$} &
  \multicolumn{1}{c}{$H_{\beta}$} &
  \multicolumn{1}{c}{$H_{\alpha}$} &
  \multicolumn{1}{c}{$H_{\alpha}$} &
  \multicolumn{1}{c}{(\ergpcmsqps)} &
  \multicolumn{1}{c}{} \\
  \multicolumn{1}{c}{(1)} &
  \multicolumn{1}{c}{(2)} &
  \multicolumn{1}{c}{(3)} &
  \multicolumn{1}{c}{(4)} &
  \multicolumn{1}{c}{(5)} &
  \multicolumn{1}{c}{(6)} &
  \multicolumn{1}{c}{(7)} &
  \multicolumn{1}{c}{(8)} &
  \multicolumn{1}{c}{(9)} &
  \multicolumn{1}{c}{(10)} &
  \multicolumn{1}{c}{(11)} &
  \multicolumn{1}{c}{(12)} &
  \multicolumn{1}{c}{(13)} &
  \multicolumn{1}{c}{(14)} &
  \multicolumn{1}{c}{(15)} \\
\hline
ESO~121-G006 & 06h07m29.5s & --61d48m28s & 14.5 & $0.33\times0.79$ & $0.78\times1.57$ & Sc & 9.70 & - & - & - & - & - & - & -\\
ESO~209-G009 & 07h58m14.9s & --49d51m09s & 11.8 & $0.27\times0.65$ & $0.64\times1.28$ & Scd & 9.81 & - & - & - & - & - & - & -\\
ESO~265-G007 & 11h07m48.8s & --46d31m38s & 11.7 & $0.27\times0.64$ & $0.63\times1.26$ & Scd & 9.55 & - & - & - & - & - & - & -\\
IC~2056 & 04h16m24.2s & --60d12m26s & 13.8 & $0.31\times0.76$ & $0.74\times1.49$ & Sbc & 9.62 & - & - & - & - & - & - & -\\
IC~342 & 03h46m49.4s & +68d05m49s & 4.6 & $0.10\times0.25$ & $0.25\times0.50$ & Scd & 10.17 & HII & 3.27 & 0.10 & 0.22 & 0.45 & 0.90 & 1\\
IIZW40 & 05h55m42.0s & +03d23m30s & 11.6 & $0.26\times0.64$ & $0.62\times1.25$ & Sbc & 9.47 & HII & 2.89 & 6.31 & 0.05 & 0.04 & 22574.40 & 6\\
M~82 & 09h55m53.1s & +69d40m41s & 3.6 & $0.08\times0.20$ & $0.19\times0.40$ & Ir & 10.77 & HII & 21.93 & 0.36 & 0.18 & 0.56 & 66.15 & 1\\
NGC~0253 & 00h47m33.1s & --25d17m15s & 3.1 & $0.07\times0.17$ & $0.17\times0.34$ & Sc & 10.44 & HII & 0.47 & 0.36 & 0.34 & 0.72 & 27.06 & 2\\
NGC~0278 & 00h52m04.3s & +47d33m01s & 11.4 & $0.26\times0.62$ & $0.61\times1.23$ & Sab & 10.03 & HII & 3.84 & 0.24 & 0.28 & 0.50 & 1.89 & 1\\
NGC~0613 & 01h34m17.8s & --29d25m10s & 15.0 & $0.34\times0.82$ & $0.81\times1.62$ & Sbc & 10.37 & HII & 6.80 & 0.48 & 0.35 & 0.59 & 17.22 & 4\\
NGC~0628 & 01h36m41.2s & +15d47m29s & 10.0 & $0.23\times0.55$ & $0.54\times1.08$ & Sc & 9.95 & HII & - & 0.61 & 0.13 & 0.41 & 7.62 & 8\\
NGC~0660 & 01h43m02.1s & +13d38m45s & 12.3 & $0.28\times0.67$ & $0.66\times1.33$ & Sa & 10.49 & L & 13.65 & 2.19 & 0.43 & 0.85 & 7.72 & 1\\
NGC~0891 & 02h22m33.5s & +42d21m18s & 8.6 & $0.20\times0.47$ & $0.46\times0.93$ & Sab & 10.27 & HII & 12.37 & 1.00 & 0.30 & 0.29 & 0.18 & 1\\
NGC~1055 & 02h41m44.3s & +00d26m36s & 11.3 & $0.26\times0.62$ & $0.61\times1.22$ & Sb & 10.09 & L & - & - & 0.41 & 0.66 & - & 1\\
NGC~1068 & 02h42m41.4s & +00d00m45s & 13.7 & $0.31\times0.75$ & $0.74\times1.48$ & Sb & 11.27 & S2 (1.8) & 5.29 & 12.82 & 0.24 & 0.76 & 38235.10 & 1\\
NGC~1448 & 03h44m32.1s & --44d38m41s & 11.5 & $0.26\times0.63$ & $0.62\times1.24$ & Scd & 9.78 & HII & $>1.67$ & - & 1.60 & 0.60 & $<0.60$ & 4\\
NGC~1559 & 04h17m37.4s & --62d46m59s & 12.7 & $0.29\times0.70$ & $0.68\times1.37$ & Scd & 10.21 & HII & $>3.14$ & $>1.00$ & 0.82 & - & 0.70 & 4\\
NGC~1569 & 04h30m49.5s & +64d51m01s & 4.6 & $0.10\times0.25$ & $0.25\times0.50$ & Ir & 9.49 & HII & 3.74 & 5.48 & 0.03 & 0.04 & 10.14 & 1\\
NGC~1792 & 05h05m13.7s & --37d58m46s & 12.5 & $0.28\times0.68$ & $0.67\times1.35$ & Sbc & 10.33 & HII & - & 0.16 & 0.21 & 0.54 & 0.30 & 4\\
NGC~1808 & 05h07m42.3s & --37d30m48s & 12.6 & $0.29\times0.69$ & $0.68\times1.36$ & Sa & 10.71 & HII & 2.95 & 0.16 & 0.21 & 0.54 & 16.31 & 2\\
NGC~2681 & 08h53m33.6s & +51d18m46s & 12.5 & $0.28\times0.68$ & $0.67\times1.35$ & Sa & 9.53 & L & 4.77 & 1.74 & 0.76 & 2.34 & 12.66 & 1\\
NGC~2903 & 09h32m10.5s & +21d30m05s & 8.3 & $0.19\times0.45$ & $0.45\times0.90$ & Sbc & 10.19 & HII & 4.42 & 0.03 & 0.19 & 0.34 & 2.66 & 1\\
NGC~3059 & 09h50m08.1s & --73d55m24s & 14.2 & $0.32\times0.78$ & $0.76\times1.54$ & Sbc & 10.00 & - & - & - & - & - & - & -\\
NGC~3175 & 10h14m42.9s & --28d52m25s & 13.4 & $0.31\times0.73$ & $0.72\times1.45$ & Sb & 9.96 & HII & - & - & - & - & - & -\\
NGC~3184 & 10h18m12.1s & +41d25m53s & 12.6 & $0.29\times0.69$ & $0.68\times1.36$ & Scd & 9.80 & HII & 4.42 & 0.13 & 0.20 & 0.33 & 1.07 & 1\\
NGC~3198 & 10h19m55.6s & +45d32m54s & 13.8 & $0.31\times0.76$ & $0.74\times1.49$ & Sc & 9.75 & HII & 6.44 & 0.23 & 0.32 & 0.42 & 0.41 & 1\\
NGC~3351 & 10h43m58.1s & +11d42m10s & 10.0 & $0.23\times0.55$ & $0.54\times1.08$ & Sb & 9.83 & HII & 4.38 & 0.27 & 0.24 & 0.46 & 2.19 & 1\\
NGC~3368 & 10h46m45.6s & +11d49m13s & 10.5 & $0.24\times0.58$ & $0.57\times1.14$ & Sab & 9.67 & L & 7.46 & 1.82 & 0.98 & 1.12 & 3.24 & 1\\
NGC~3511 & 11h03m24.2s & --23d05m15s & 13.6 & $0.31\times0.75$ & $0.73\times1.47$ & Sc & 9.82 & HII & 12.30 & 0.28 & 0.27 & 0.44 & - & 7\\
NGC~3521 & 11h05m49.2s & +00d02m15s & 6.8 & $0.15\times0.37$ & $0.37\times0.74$ & Sbc & 9.96 & L & 3.71 & 1.00 & 0.87 & 0.65 & 7.39 & 1\\
NGC~3556 & 11h11m31.5s & +55d40m23s & 13.9 & $0.32\times0.76$ & $0.75\times1.50$ & Scd & 10.37 & HII & 7.29 & 0.26 & 0.29 & 0.32 & 1.81 & 1\\
NGC~3621 & 11h18m16.7s & --32d48m47s & 6.6 & $0.15\times0.36$ & $0.36\times0.71$ & Sd & 9.74 & S2 & 3.42 & 8.61 & 0.21 & 1.46 & 2.24 & 9\\
NGC~3627 & 11h20m15.3s & +12d59m32s & 10.0 & $0.23\times0.55$ & $0.54\times1.08$ & Sb & 10.38 & L & 5.92 & 2.90 & 0.76 & 1.45 & 12.95 & 1\\
NGC~3628 & 11h20m17.4s & +13d35m19s & 10.0 & $0.23\times0.55$ & $0.54\times1.08$ & Sb & 10.25 & L & 4.72 & 1.78 & 0.78 & 0.95 & 0.23 & 1\\
NGC~3675 & 11h26m09.0s & +43d35m04s & 12.7 & $0.29\times0.70$ & $0.68\times1.37$ & Sab & 9.92 & L & 3.20 & 1.29 & 0.68 & 1.48 & 3.21 & 1\\
NGC~3726 & 11h33m19.9s & +47d01m49s & 14.5 & $0.33\times0.79$ & $0.78\times1.57$ & Sc & 9.78 & HII & 3.44 & 0.14 & 0.22 & 0.31 & 2.52 & 1\\
NGC~3938 & 11h52m49.2s & +44d07m14s & 14.8 & $0.34\times0.81$ & $0.80\times1.60$ & Sc & 9.93 & HII & 2.82 & 1.78 & 0.83 & 0.52 & 1.19 & 1\\
NGC~3949 & 11h53m42.0s & +47d51m31s & 13.6 & $0.31\times0.75$ & $0.73\times1.47$ & Sbc & 9.87 & HII & 3.51 & 0.22 & 0.44 & 0.40 & 1.24 & 1\\
NGC~4013 & 11h58m31.5s & +43d56m54s & 13.8 & $0.31\times0.76$ & $0.74\times1.49$ & Sb & 9.76 & L & 2.05 & 0.71 & 0.83 & 1.12 & 1.01 & 1\\
NGC~4051 & 12h03m09.8s & +44d31m50s & 13.1 & $0.30\times0.72$ & $0.70\times1.42$ & Sbc & 9.90 & S2 (1.5) & 3.30 & 4.47 & 0.36 & 0.65 & 726.25 & 1\\
\hline\end{tabular}
\end{center}
\end{minipage}
\end{table}  
\end{landscape}

\begin{landscape}
\begin{table}
\begin{minipage}{240mm}
\begin{center}
\vspace{2.0cm}
\contcaption{}
\begin{tabular}{lccrcccrccccccc}
\hline
  \multicolumn{1}{c}{Common} &
  \multicolumn{1}{c}{RA} &
  \multicolumn{1}{c}{DEC} &
  \multicolumn{1}{c}{Dist} &
  \multicolumn{1}{c}{SH} &
  \multicolumn{1}{c}{LH} &
  \multicolumn{1}{c}{Morph} &
  \multicolumn{1}{c}{$\Lir$} &
  \multicolumn{1}{c}{Spectral} &
  \multicolumn{1}{c}{\underline{$H_{\alpha}$}} &
  \multicolumn{1}{c}{\underline{[OIII]}} &
  \multicolumn{1}{c}{\underline{[SII]}} &
  \multicolumn{1}{c}{\underline{[NII]}} &
  \multicolumn{1}{c}{$10^{15} f_{\rm{[OIII]}}$} &
  \multicolumn{1}{c}{Ref.} \\
  \multicolumn{1}{c}{Name} &
  \multicolumn{1}{c}{} &
  \multicolumn{1}{c}{} &
  \multicolumn{1}{c}{(Mpc)} &
  \multicolumn{1}{c}{(kpc)} &
  \multicolumn{1}{c}{(kpc)} &
  \multicolumn{1}{c}{Class} &
  \multicolumn{1}{c}{($L_{\odot}$)} &
  \multicolumn{1}{c}{Class} &
  \multicolumn{1}{c}{$H_{\beta}$} &
  \multicolumn{1}{c}{$H_{\beta}$} &
  \multicolumn{1}{c}{$H_{\alpha}$} &
  \multicolumn{1}{c}{$H_{\alpha}$} &
  \multicolumn{1}{c}{(\ergpcmsqps)} &
  \multicolumn{1}{c}{} \\
  \multicolumn{1}{c}{(1)} &
  \multicolumn{1}{c}{(2)} &
  \multicolumn{1}{c}{(3)} &
  \multicolumn{1}{c}{(4)} &
  \multicolumn{1}{c}{(5)} &
  \multicolumn{1}{c}{(6)} &
  \multicolumn{1}{c}{(7)} &
  \multicolumn{1}{c}{(8)} &
  \multicolumn{1}{c}{(9)} &
  \multicolumn{1}{c}{(10)} &
  \multicolumn{1}{c}{(11)} &
  \multicolumn{1}{c}{(12)} &
  \multicolumn{1}{c}{(13)} &
  \multicolumn{1}{c}{(14)} &
  \multicolumn{1}{c}{(15)} \\
\hline
NGC~4085 & 12h05m23.3s & +50d21m08s & 14.6 & $0.33\times0.80$ & $0.79\times1.58$ & Sc & 9.66 & HII & 4.71 & 0.63 & 0.19 & 0.32 & 172.71 & 6\\
NGC~4088 & 12h05m35.1s & +50d32m24s & 13.4 & $0.31\times0.73$ & $0.72\times1.45$ & Sbc & 10.25 & HII & 7.04 & 0.21 & 0.18 & 0.32 & 1.40 & 1\\
NGC~4157 & 12h11m04.2s & +50d29m04s & 13.3 & $0.30\times0.73$ & $0.72\times1.44$ & Sab & 10.12 & HII & 10.69 & 0.22 & 0.24 & 0.34 & 0.43 & 1\\
NGC~4490 & 12h30m34.9s & +41d38m47s & 10.5 & $0.24\times0.58$ & $0.57\times1.14$ & Sd & 10.21 & L & 6.39 & 2.57 & 0.71 & 0.25 & 1.83 & 1\\
NGC~4536 & 12h34m28.5s & +02d11m08s & 14.9 & $0.34\times0.82$ & $0.80\times1.61$ & Sbc & 10.32 & HII & 5.23 & 0.33 & 0.36 & 0.47 & 5.42 & 1\\
NGC~4559 & 12h35m57.0s & +27d57m37s & 9.7 & $0.22\times0.53$ & $0.52\times1.05$ & Scd & 9.56 & HII & 3.59 & 0.35 & 0.40 & 0.42 & 0.90 & 1\\
NGC~4631 & 12h42m07.1s & +32d32m33s & 7.7 & $0.18\times0.42$ & $0.41\times0.83$ & Sd & 10.22 & HII & 3.07 & 1.51 & 0.23 & 0.24 & 1.85 & 1\\
NGC~4666 & 12h45m07.7s & +00d27m41s & 12.8 & $0.29\times0.70$ & $0.69\times1.38$ & Sc & 10.36 & L & 5.13 & 1.20 & 0.60 & 1.29 & 558.28 & 3\\
NGC~4736 & 12h50m52.9s & +41d07m15s & 4.8 & $0.11\times0.26$ & $0.26\times0.52$ & Sab & 9.73 & L & 3.14 & 1.47 & 1.39 & 2.15 & 9.59 & 1\\
NGC~4818 & 12h56m50.0s & --08d31m38s & 9.4 & $0.21\times0.51$ & $0.51\times1.02$ & Sab & 9.75 & HII & 2.95 & 0.15 & 0.18 & 0.63 & - & 2\\
NGC~4945 & 13h05m27.6s & --49d28m09s & 3.9 & $0.09\times0.21$ & $0.21\times0.42$ & Scd & 10.48 & HII & 1.20 & $<0.90$ & 0.15 & 0.30 & $<1.00$ & 5\\
NGC~5033 & 13h13m27.2s & +36d35m40s & 13.8 & $0.31\times0.76$ & $0.74\times1.49$ & Sc & 10.13 & S2 (1.9) & 4.48 & 4.68 & 1.07 & 2.34 & 96.28 & 1\\
NGC~5055 & 13h15m49.5s & +42d01m39s & 8.0 & $0.18\times0.44$ & $0.43\times0.86$ & Sbc & 10.09 & L & 5.42 & 1.86 & 0.74 & 1.48 & 3.61 & 1\\
NGC~5128 & 13h25m27.6s & --43d01m12s & 4.0 & $0.09\times0.22$ & $0.22\times0.43$ & S0 & 10.11 & HII & 5.50 & 0.55 & 0.28 & 0.45 & 12.06 & 4\\
NGC~5194 & 13h29m53.5s & +47d11m42s & 8.6 & $0.20\times0.47$ & $0.46\times0.93$ & Sbc & 10.42 & S2 & 8.44 & 8.91 & 0.85 & 2.88 & 90.87 & 1\\
NGC~5195 & 13h30m00.0s & +47d16m00s & 8.3 & $0.19\times0.45$ & $0.45\times0.90$ & Irr & 9.50 & L & 1.90 & 1.22 & 2.00 & 5.37 & 6.76 & 1\\
NGC~5236 & 13h36m58.8s & --29d51m46s & 3.6 & $0.08\times0.20$ & $0.19\times0.39$ & Sc & 10.10 & HII & 6.05 & 0.29 & 0.21 & 1.33 & 35.36 & 4\\
NGC~5643 & 14h32m41.1s & --44d10m30s & 13.9 & $0.32\times0.76$ & $0.75\times1.50$ & Sc & 10.24 & S2 & 7.80 & 10.67 & 0.71 & 1.15 & 607.36 & 4\\
NGC~5907 & 15h15m58.9s & +56d18m36s & 12.1 & $0.28\times0.66$ & $0.65\times1.31$ & Sc & 9.85 & HII & 11.81 & 1.07 & 0.34 & 0.60 & 0.44 & 1\\
NGC~6300 & 17h17m00.3s & --62d49m13s & 13.1 & $0.30\times0.72$ & $0.70\times1.42$ & Sb & 10.09 & S2 & 7.44 & 15.14 & 0.14 & 0.95 & 38.69 & 4\\
NGC~6744 & 19h09m45.9s & --63d51m27s & 9.9 & $0.23\times0.54$ & $0.53\times1.07$ & Sbc & 9.99 & L & 0.90 & 1.30 & 2.44 & - & 6.75 & 4\\
NGC~6946 & 20h34m52.6s & +60d09m12s & 5.3 & $0.12\times0.29$ & $0.29\times0.57$ & Scd & 10.16 & HII & 9.03 & 0.38 & 0.32 & 0.64 & 3.17 & 1\\
NGC~7331 & 22h37m04.6s & +34d24m56s & 14.7 & $0.33\times0.81$ & $0.79\times1.59$ & Sb & 10.58 & L & 3.92 & 2.79 & 0.55 & 1.44 & 7.77 & 1\\
UGCA~127 & 06h20m56.9s & --08d29m42s & 10.2 & $0.23\times0.56$ & $0.55\times1.10$ & Scd & 9.82 & - & - & - & - & - & - & -\\
\hline\end{tabular}
\end{center}
\medskip
{\tiny NOTES:} (1) Common galaxy name.  (2--3) {\it 2MASS} near-IR
position of galactic nucleus.  (4) Luminosity distance to source in
Mpc from the RBGS (Sanders et al. 2003).  (5--6) Projected spectral
apertures (SH and LH respectively) in kiloparsecs. (7) Morphological
classification from RC3 \citep{rc3}.  (8) Logarithm of IR luminosity
(8--1000$\um$) from RBGS.
(9) Optical spectral class from BPT diagnostics (S2: Seyfert 2;
broad-line sub-class in parentheses, HII: Star-forming galaxy, and L:
LINER).  (10--13) Optical emission line ratios. (14) Observed [OIII]
flux in units of $10^{-15} \ergpcmsqps$.  (15) References for
published optical data.

{\tiny REFERENCES:} (1) \citet{ho97a}; (2) \citet{kewley01}; 
(3) \citet{veilleux95}; (4) Veron-Cetty et al. (1986); (5)
Moorwood et al. (1996); (6) \citet{moustakas06}; (7)
\citet{kirhakos90}; (8) \citet{ganda06}; (9) \citet{barth09}.
\end{minipage}
\end{table}
\end{landscape}


\noindent distance of $D < 15$ Mpc with $\Lir > 3 \times 10^9 \Lsun$,
64 of which have \spitzer-IRS high-resolution spectroscopy publicly
available (i.e., $\approx 94$ percent complete): P3124 (28 objects;
PI: D.M.~Alexander), P159 (18 objects; PI: R.~Kennicutt [SINGS]), P14
(11 objects; PI: J.R.~Houck), P59 (4 objects; PI: G.~Rieke) and P86 (3
objects; PI: M.~Werner).

In Fig.~\ref{fig_1} we plot IR luminosity versus luminosity distance
for the RBGS and highlight the 64 galaxies with {\it Spitzer}-IRS
observations in our $D < 15$~Mpc sample. The basic properties from the
RBGS for the sources are combined with published optical data and
listed in Table 1. The objects are all late-type galaxies (Hubble
classification of S0 or later), which is unsurprising since early-type
galaxies are typically IR faint and undetected by {\it IRAS} (e.g.,\
\citealt{knapp89}). The four galaxies that match our selection
criteria but lack sufficient high-resolution \spitzer-IRS observations
of the central regions are shown in Table 2. Specifically, NGC~3486
has Short-High (SH) and Long-High (LH) observations but the data are
noisy and no statistically useful information can be extracted;
NGC~4565 and NGC~5457 have only high-resolution observations of
extranuclear regions; NGC~5248 has no high-resolution
observations. NGC~5248 and NGC~5457 are optically classified as
star-forming HII galaxies while NGC~3486 and NGC~4565 are optically
classified as Seyfert galaxies (Ho97). Both NGC~3486 and NGC~4565 have
published \oiv fluxes from low-resolution ($R \sim 100$) \spitzer-IRS
spectroscopy in \citet{diamond09}; the derived \oiv luminosities
suggest that the AGNs are contributing $< 5$ percent to the total
bolometric luminosity of the galaxy.


\begin{table}
\caption{$D<15$ Mpc IR-bright galaxies not included in sample}
\begin{center}
\setlength{\tabcolsep}{0.6mm}
\begin{tabular}{lcccccc}

\hline
  \multicolumn{1}{c}{Common} &
  \multicolumn{1}{c}{Morph} &
  \multicolumn{1}{c}{Dist} &
  \multicolumn{1}{c}{$\Lir$} &
  \multicolumn{1}{c}{Spectral} &
  \multicolumn{1}{c}{\spitzer-IRS} &
  \multicolumn{1}{c}{L$_{\rm [OIV]}$} \\
  \multicolumn{1}{c}{Name} &
  \multicolumn{1}{c|}{Class} &
  \multicolumn{1}{c}{(Mpc)} &
  \multicolumn{1}{c}{(${\rm L}_{\odot}$)} &
  \multicolumn{1}{c}{Class} &
  \multicolumn{1}{c}{Spec?} &
  \multicolumn{1}{c}{(\ergps)} \\
  \multicolumn{1}{c}{(1)} &
  \multicolumn{1}{c}{(2)} &
  \multicolumn{1}{c}{(3)} &
  \multicolumn{1}{c}{(4)} &
  \multicolumn{1}{c}{(5)} &
  \multicolumn{1}{c}{(6)} &
  \multicolumn{1}{c}{(7)} \\
\hline
NGC~3486 & Sc & 9.2 & 9.31 & Sy2 & 1 & $38.52$ \\
NGC~4565 & Sb & 10.0 & 9.66 & Sy2 & 2 & $38.40$ \\
NGC~5248 & Sbc & 13.8 & 10.20 & HII & 3 & - \\
NGC~5457 & Scd & 6.7 & 10.20 & HII & 2 & - \\
\hline
\end{tabular}
\end{center}
\medskip
{\tiny NOTES:} (1) Common galaxy name. (2) Morphological
classification from RC3. (3) Luminosity distance in Mpc from RBGS. (4)
Logarithm of IR luminosity (8--1000$\um$) from RBGS. (5) Optical
spectral class from Ho97 using BPT diagnostics; see Table~1. (6)
Status of \spitzer-IRS data: 1. High-resolution observations are
publicly available but signal-to-noise is not sufficient for analysis;
2. High-resolution observations available only for extranuclear
regions; 3. No high-resolution observations available. (7) Logarithm
of \oiv luminosity from low-resolution ($R \sim 100$) \spitzer-IRS
spectroscopy in units of \ergps \ (Diamond-Stanic et al. 2009).
\label{tbl_missing}
\end{table}

\begin{figure}
\includegraphics[width=\linewidth]{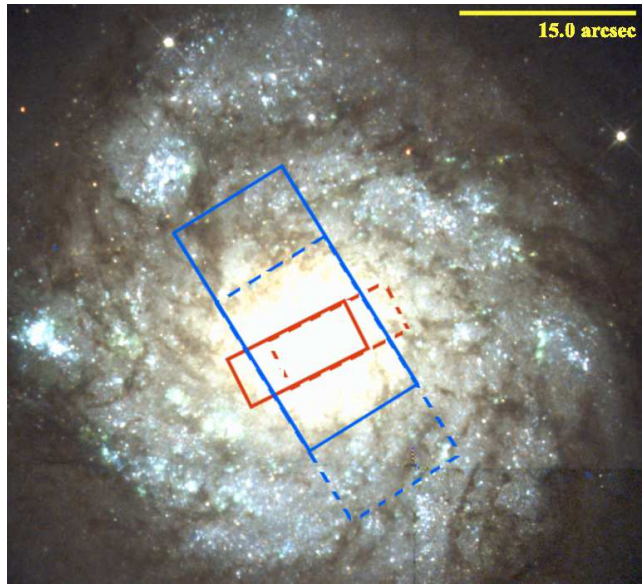}
\caption{{\it Hubble Space Telescope} ({\it HST}) WFPC2 (3-colour)
optical image of the circumnuclear region of a typical galaxy in the
$D<15$ Mpc sample, with \spitzer-IRS SH ($4.7" \times 11.3"$, $\lambda
\sim 9.9$--$19.6 \um$) and LH ($11.1" \times 22.3"$, $\lambda \sim
18.7$--$37.2 \um$) apertures overlaid in the two nod positions
(staring-mode observations). The object shown is the Sbc galaxy,
NGC~0278 at $D \approx 11.4$ Mpc with $\Lir \approx 1.1 \times 10^{10}
\Lsun$. Linear size-scales of apertures are $\approx 0.26 \times 0.63$
kpc (SH) and $\approx 0.61 \times 1.23$ kpc (LH). In highly resolved
sources such as NGC~0278 the two nod positions will produce slightly
differing spectra but the inner central region will be bright and thus
will dominate.}
\label{fig_overlay}
\end{figure}

\subsection{Data Reduction}
\label{sec:data_red}
Each of the galaxies in our $D<15$~Mpc sample were observed using both
the SH ($4.7" \times 11.3"$, $\lambda \sim 9.9$--$19.6 \um$) and LH
($11.1" \times 22.3"$, $\lambda \sim 18.7$--$37.2 \um$) resolution
spectrographs onboard the NASA \spitzer \ Space Telescope
(\citealt{spitzer_irs}; \citealt{spitzer_mission}). The spectral
resolution is $R \sim 600$ for both SH and LH modules. The raw data
are compiled from multiple observing programs (see \S2.1) and they
consist of both spectral mapping and staring observations, with
differing exposure times. As an example, in Fig. \ref{fig_overlay} we
project the SH and LH apertures onto a galaxy in our sample
(NGC~0278), showing the two differing nod positions.

Many of the objects in our sample do not have dedicated off-source
observations. However, as we only require emission-line flux
measurements no background subtraction was necessary for the
observations (see \S7.2.5.2 of the \spitzer-IRS Observers
Manual).\footnote{The \spitzer-IRS Observers Manual is available at
http://ssc.spitzer.caltech.edu/irs/dh/} Furthermore, the targets are
bright compared to the background and therefore any background
corrections would be small.

For the reduction of the IRS-staring data, Basic Calibrated Data (BCD)
images were co-added and mean averaged at each nod position. Few of
the observations were pre-processed with the same \spitzer \ pipeline
version. Therefore, to ensure consistency, we extracted the spectra of
each galaxy using a custom pipeline based on the \spitzer \ data
reduction packages {\small IRSCLEAN} (to apply individual custom
bad-pixel masks for each of the BCDs) and {\small SPICE} (to extract
full slit spectra using the latest flux calibration files: version
17.2). For further information on the high-resolution \spitzer-IRS
pre-processing pipeline, see Chapter 7 of the \spitzer \ Observers
Manual.


\begin{landscape}
\begin{table}
\begin{minipage}{240mm}
\begin{center}
\setlength{\tabcolsep}{1.5mm}
\vspace{1.0cm}
\caption{Mid-IR Spectral Emission Lines ($10^{-14}$ erg s$^{-1}$ cm$^{-2}$)}
\begin{tabular}{lcccccccccc}
\hline
  \multicolumn{1}{c}{Common} &
  \multicolumn{1}{c}{EW PAH} &
  \multicolumn{1}{c}{[NeII]} &
  \multicolumn{1}{c}{[NeV]} &
  \multicolumn{1}{c}{[NeIII]} &
  \multicolumn{1}{c}{[NeV]} &
  \multicolumn{1}{c}{[OIV]} &
  \multicolumn{1}{c}{[FeII]} &
  \multicolumn{1}{c}{[SIII]} &
  \multicolumn{1}{c}{[SiII]} &
  \multicolumn{1}{c}{Mid-IR} \\
  \multicolumn{1}{c}{Name} &
  \multicolumn{1}{c}{$\lambda 11.3 \um$} &
  \multicolumn{1}{c}{$\lambda 12.81 \um$} &
  \multicolumn{1}{c}{$\lambda 14.32 \um$} &
  \multicolumn{1}{c}{$\lambda 15.56 \um$} &
  \multicolumn{1}{c}{$\lambda 24.32 \um$} &
  \multicolumn{1}{c}{$\lambda 25.89 \um$} &
  \multicolumn{1}{c}{$\lambda 25.99 \um$} &
  \multicolumn{1}{c}{$\lambda 33.48 \um$} &
  \multicolumn{1}{c}{$\lambda 34.82 \um$} &
  \multicolumn{1}{c}{AGN?} \\
  \multicolumn{1}{c}{} &
  \multicolumn{1}{c}{($\umu {\rm m}$)} &
  \multicolumn{1}{c}{21.6 eV} &
  \multicolumn{1}{c}{97.1 eV} &
  \multicolumn{1}{c}{41.0 eV} &
  \multicolumn{1}{c}{97.1 eV} &
  \multicolumn{1}{c}{54.9 eV} &
  \multicolumn{1}{c}{7.9 eV} &
  \multicolumn{1}{c}{23.3 eV} &
  \multicolumn{1}{c}{8.2 eV} &
  \multicolumn{1}{c}{} \\
  \multicolumn{1}{c}{(1)} &
  \multicolumn{1}{c}{(2)} &
  \multicolumn{1}{c}{(3)} &
  \multicolumn{1}{c}{(4)} &
  \multicolumn{1}{c}{(5)} &
  \multicolumn{1}{c}{(6)} &
  \multicolumn{1}{c}{(7)} &
  \multicolumn{1}{c}{(8)} &
  \multicolumn{1}{c}{(9)} &
  \multicolumn{1}{c}{(10)} &
  \multicolumn{1}{c}{(11)} \\
\hline
  ESO~121-G006 & 0.816 & $16.48\pm0.50$ & $0.65\pm0.08$ & $3.24\pm0.24$ & $<0.59$ & $4.57\pm0.40$ & $3.03\pm0.26$ & $28.59\pm2.80$ & $45.88\pm1.87$ & Y \\
  ESO~209-G009 & 1.127 & $9.81\pm0.59$ & $<0.26$ & $1.43\pm0.20$ & $<0.12$ & $<0.29$ & $<0.99$ & $22.86\pm2.53$ & $31.56\pm1.70$ & - \\
  ESO~265-G007 & 19.155 & $2.62\pm0.32$ & $<0.41$ & $1.13\pm0.23$ & $<1.18$ & $<1.04$ & $1.59\pm0.17$ & $8.47\pm0.92$ & $11.83\pm1.08$ & - \\
  IC~2056 & 1.072 & $12.77\pm0.26$ & $<0.16$ & $2.15\pm0.15$ & $<0.10$ & $0.75\pm0.16$ & $2.95\pm0.21$ & $42.00\pm6.98$ & $52.60\pm2.00$ & - \\
  IC~342 & 0.498 & $669.38\pm5.01$ & $<0.64$ & $39.21\pm0.73$ & $<13.71$ & $<6.83$ & $60.45\pm4.95$ & $652.93\pm18.77$ & $953.15\pm13.71$ & - \\
  IIZW40 & 0.015 & $6.04\pm1.12$ & $<0.28$ & $117.34\pm2.34$ & $<1.14$ & $8.49\pm2.28$ & $1.92\pm0.49$ & $86.92\pm2.02$ & $36.05\pm1.41$ & - \\
  M~82 & 0.843 & $506.22\pm45.04$ & $<0.89$ & $80.98\pm1.99$ & $<14.65$ & $45.67\pm6.61$ & $137.77\pm0.66$ & $1812.90\pm36.79$ & $2166.00\pm25.52$ & - \\
  NGC~0253 & 0.295 & $3099.10\pm126.60$ & $<5.13$ & $207.89\pm9.05$ & $<113.68$ & $<83.66$ & $245.20\pm0.93$ & $1463.40\pm69.65$ & $2411.80\pm64.02$ & - \\
  NGC~0278 & 0.851 & $17.84\pm0.37$ & $<0.07$ & $2.95\pm0.16$ & $<0.59$ & $<0.62$ & $3.82\pm0.64$ & $49.55\pm3.37$ & $86.82\pm1.07$ & - \\
  NGC~0613 & 0.501 & $130.73\pm1.55$ & $0.67\pm0.11$ & $15.96\pm0.50$ & $3.18\pm0.56$ & $9.09\pm1.53$ & $12.47\pm1.25$ & $84.23\pm2.70$ & $162.49\pm2.11$ & Y \\
  NGC~0628 & 0.175 & $2.31\pm0.36$ & $<0.15$ & $<0.08$ & $<0.46$ & $<0.13$ & $<0.11$ & $12.35\pm0.94$ & $4.78\pm0.40$ & - \\
  NGC~0660 & 0.730 & $379.81\pm13.96$ & $3.88\pm0.25$ & $37.46\pm0.69$ & $<4.59$ & $28.26\pm8.24$ & $22.03\pm3.61$ & $250.60\pm11.47$ & $435.29\pm4.03$ & Y \\
  NGC~0891 & 0.364 & $8.57\pm0.78$ & $<0.04$ & $0.84\pm0.07$ & $<0.62$ & $<1.03$ & $<0.84$ & $10.74\pm2.02$ & $28.11\pm0.92$ & - \\
  NGC~1055 & 0.747 & $26.42\pm0.79$ & $<0.19$ & $2.58\pm0.20$ & $<0.58$ & $1.22\pm0.26$ & $3.32\pm0.47$ & $28.26\pm2.46$ & $71.16\pm0.98$ & - \\
  NGC~1068 & 0.008 & $538.34\pm37.3$ & $898.04\pm42.25$ & $1432.20\pm76.87$ & $815.96\pm58.78$ & $2066.90\pm154.88$ & $169.76\pm41.70$ & $378.51\pm34.07$ & $616.32\pm21.94$ & Y \\
  NGC~1448 & 0.378 & $8.17\pm0.72$ & $3.68\pm0.29$ & $5.02\pm0.12$ & $3.60\pm0.62$ & $16.02\pm0.49$ & $1.60\pm0.26$ & $23.76\pm5.08$ & $22.55\pm1.13$ & Y \\
  NGC~1559 & 0.652 & $14.88\pm0.24$ & $<0.44$ & $2.34\pm0.18$ & $<0.36$ & $<0.69$ & $2.91\pm0.21$ & $23.25\pm4.32$ & $26.00\pm0.88$ & - \\
  NGC~1569 & 0.089 & $19.06\pm1.28$ & $<0.38$ & $188.15\pm2.69$ & $<1.37$ & $29.32\pm0.16$ & $6.43\pm0.27$ & $188.06\pm4.38$ & $111.08\pm2.59$ & - \\
  NGC~1792 & 0.492 & $23.32\pm0.32$ & $0.41\pm0.13$ & $2.10\pm0.17$ & $<0.25$ & $0.96\pm0.06$ & $3.38\pm0.38$ & $31.01\pm0.93$ & $57.43\pm0.72$ & Y \\
  NGC~1808 & 0.682 & $177.36\pm16.36$ & $<0.91$ & $17.26\pm0.68$ & $<8.38$ & $<9.54$ & $16.62\pm2.78$ & $205.83\pm20.27$ & $354.26\pm15.15$ & - \\
  NGC~2681 & 0.883 & $6.88\pm0.28$ & $<0.15$ & $3.43\pm0.21$ & $<0.36$ & $2.17\pm0.70$ & $1.56\pm0.43$ & $<4.01$ & $7.77\pm0.66$ & - \\
  NGC~2903 & 0.699 & $181.62\pm13.59$ & $<0.82$ & $13.52\pm0.18$ & $<0.94$ & $<1.70$ & $13.38\pm2.73$ & $124.25\pm1.85$ & $298.89\pm2.03$ & - \\
  NGC~3059 & 0.882 & $26.79\pm0.93$ & $<0.18$ & $3.13\pm0.25$ & $<0.69$ & $<1.48$ & $4.07\pm0.64$ & $40.41\pm5.92$ & $59.18\pm0.82$ & - \\
  NGC~3175 & 0.752 & $34.08\pm0.46$ & $<0.35$ & $2.46\pm0.22$ & $<0.86$ & $1.15\pm0.07$ & $3.91\pm1.54$ & $71.51\pm2.47$ & $92.40\pm1.28$ & - \\
  NGC~3184 & 0.324 & $5.70\pm0.15$ & $<0.10$ & $0.67\pm0.08$ & $<0.20$ & $0.50\pm0.09$ & $<0.27$ & $4.31\pm0.35$ & $5.05\pm0.52$ & - \\
  NGC~3198 & 0.423 & $4.75\pm0.11$ & $<0.03$ & $0.30\pm0.08$ & $<0.16$ & $0.45\pm0.10$ & $<0.31$ & $3.39\pm0.45$ & $4.34\pm0.59$ & - \\
  NGC~3351 & 0.263 & $18.87\pm1.09$ & $<0.13$ & $1.71\pm0.08$ & $<0.14$ & $4.71\pm3.19$ & $2.91\pm0.32$ & $38.47\pm1.41$ & $70.25\pm1.19$ & - \\
  NGC~3368 & 0.488 & $4.29\pm0.29$ & $<0.29$ & $2.85\pm0.25$ & $<0.74$ & $0.99\pm0.29$ & $2.58\pm0.12$ & $<5.88$ & $13.22\pm2.53$ & - \\
  NGC~3511 & 0.647 & $6.37\pm0.18$ & $<0.34$ & $1.39\pm0.11$ & $<0.64$ & $<1.11$ & $<1.25$ & $20.19\pm3.06$ & $31.35\pm0.79$ & - \\
  NGC~3521 & 0.172 & $3.02\pm0.22$ & $<0.11$ & $1.77\pm0.17$ & $<0.31$ & $1.53\pm0.11$ & $<0.82$ & $7.14\pm1.08$ & $21.79\pm1.11$ & - \\
  NGC~3556 & 0.771 & $26.43\pm1.79$ & $<0.26$ & $3.44\pm0.11$ & $<0.32$ & $<0.69$ & $2.79\pm0.18$ & $100.28\pm1.09$ & $95.50\pm1.14$ & - \\
  NGC~3621 & 0.555 & $3.69\pm0.16$ & $0.29\pm0.00$ & $1.51\pm0.16$ & $0.53\pm0.05$ & $2.86\pm0.39$ & $0.86\pm0.73$ & $12.27\pm0.62$ & $17.66\pm1.08$ & Y \\
  NGC~3627 & 0.579 & $8.14\pm0.29$ & $0.30\pm0.08$ & $2.79\pm0.15$ & $0.47\pm0.04$ & $1.99\pm0.53$ & $1.30\pm0.28$ & $5.78\pm0.96$ & $16.84\pm1.90$ & Y \\
  NGC~3628 & 0.798 & $155.28\pm8.38$ & $0.95\pm0.06$ & $10.38\pm0.19$ & $<2.39$ & $5.35\pm2.89$ & $12.12\pm0.27$ & $155.61\pm6.44$ & $264.07\pm2.31$ & Y \\
  NGC~3675 & 0.588 & $8.07\pm0.13$ & $<0.23$ & $2.21\pm0.20$ & $<0.48$ & $<0.58$ & $<1.39$ & $12.65\pm3.29$ & $25.12\pm0.89$ & - \\
  NGC~3726 & 1.408 & $5.47\pm0.13$ & $<0.41$ & $0.43\pm0.06$ & $<0.52$ & $<0.44$ & $<0.45$ & $12.60\pm0.11$ & $11.82\pm0.27$ & - \\
  NGC~3938 & 0.314 & $1.43\pm0.14$ & $<0.09$ & $0.31\pm0.09$ & $<0.14$ & $<0.29$ & $<0.17$ & $3.20\pm0.29$ & $9.84\pm2.57$ & - \\
  NGC~3949 & 0.528 & $5.82\pm0.29$ & $<0.11$ & $1.11\pm0.15$ & $<0.11$ & $1.48\pm0.67$ & $1.15\pm0.03$ & $17.12\pm1.28$ & $29.25\pm0.89$ & - \\
  NGC~4013 & 0.639 & $13.12\pm0.26$ & $<0.09$ & $2.77\pm0.24$ & $<0.89$ & $<1.01$ & $<2.44$ & $13.49\pm2.33$ & $34.23\pm0.90$ & - \\
  NGC~4051 & 0.085 & $16.78\pm0.48$ & $11.26\pm0.43$ & $16.02\pm0.47$ & $11.82\pm1.20$ & $36.42\pm2.41$ & $<1.93$ & $13.74\pm1.17$ & $17.17\pm1.73$ & Y \\
 \hline\end{tabular}
 \end{center}

 \end{minipage}
 \end{table}
 \end{landscape}
  
\begin{landscape}
\begin{table}
\begin{minipage}{220mm}
\begin{center}
\setlength{\tabcolsep}{3mm}
\vspace{2.0cm}
\contcaption{}
\begin{tabular}{lcccccccccc}
\hline
  \multicolumn{1}{c}{Common} &
  \multicolumn{1}{c}{EW PAH} &
  \multicolumn{1}{c}{[NeII]} &
  \multicolumn{1}{c}{[NeV]} &
  \multicolumn{1}{c}{[NeIII]} &
  \multicolumn{1}{c}{[NeV]} &
  \multicolumn{1}{c}{[OIV]} &
  \multicolumn{1}{c}{[FeII]} &
  \multicolumn{1}{c}{[SIII]} &
  \multicolumn{1}{c}{[SiII]} &
  \multicolumn{1}{c}{Mid-IR} \\
  \multicolumn{1}{c}{Name} &
  \multicolumn{1}{c}{$\lambda 11.3 \um$} &
  \multicolumn{1}{c}{$\lambda 12.81 \um$} &
  \multicolumn{1}{c}{$\lambda 14.32 \um$} &
  \multicolumn{1}{c}{$\lambda 15.56 \um$} &
  \multicolumn{1}{c}{$\lambda 24.32 \um$} &
  \multicolumn{1}{c}{$\lambda 25.89 \um$} &
  \multicolumn{1}{c}{$\lambda 25.99 \um$} &
  \multicolumn{1}{c}{$\lambda 33.48 \um$} &
  \multicolumn{1}{c}{$\lambda 34.82 \um$} &
  \multicolumn{1}{c}{AGN?} \\
  \multicolumn{1}{c}{} &
  \multicolumn{1}{c}{($\umu {\rm m}$)} &
  \multicolumn{1}{c}{21.6 eV} &
  \multicolumn{1}{c}{97.1 eV} &
  \multicolumn{1}{c}{41.0 eV} &
  \multicolumn{1}{c}{97.1 eV} &
  \multicolumn{1}{c}{54.9 eV} &
  \multicolumn{1}{c}{7.9 eV} &
  \multicolumn{1}{c}{23.3 eV} &
  \multicolumn{1}{c}{8.2 eV} &
  \multicolumn{1}{c}{} \\
  \multicolumn{1}{c}{(1)} &
  \multicolumn{1}{c}{(2)} &
  \multicolumn{1}{c}{(3)} &
  \multicolumn{1}{c}{(4)} &
  \multicolumn{1}{c}{(5)} &
  \multicolumn{1}{c}{(6)} &
  \multicolumn{1}{c}{(7)} &
  \multicolumn{1}{c}{(8)} &
  \multicolumn{1}{c}{(9)} &
  \multicolumn{1}{c}{(10)} &
  \multicolumn{1}{c}{(11)} \\
\hline
  NGC~4085 & 1.159 & $23.03\pm0.31$ & $<0.15$ & $2.92\pm0.21$ & $<0.92$ & $<0.44$ & $1.91\pm0.43$ & $36.63\pm2.09$ & $53.22\pm1.66$ & - \\
  NGC~4088 & 0.575 & $43.67\pm1.61$ & $<0.19$ & $2.54\pm0.13$ & $<0.47$ & $<0.53$ & $1.69\pm0.31$ & $36.31\pm0.80$ & $61.27\pm0.59$ & - \\
  NGC~4157 & 0.832 & $11.63\pm0.18$ & $<0.07$ & $1.44\pm0.12$ & $<0.70$ & $1.10\pm0.32$ & $2.51\pm0.75$ & $29.70\pm3.13$ & $54.72\pm2.79$ & - \\
  NGC~4490 & 0.396 & $3.55\pm0.45$ & $<0.10$ & $3.50\pm0.12$ & $<0.10$ & $1.08\pm0.17$ & $1.75\pm0.35$ & $16.87\pm2.97$ & $22.26\pm0.87$ & - \\
  NGC~4536 & 0.434 & $35.46\pm0.39$ & $<0.06$ & $6.11\pm0.06$ & $<0.13$ & $1.72\pm0.25$ & $9.74\pm1.50$ & $100.48\pm4.53$ & $114.02\pm1.53$ & - \\
  NGC~4559 & 0.214 & $1.89\pm0.13$ & $<0.05$ & $0.53\pm0.08$ & $<0.07$ & $<1.62$ & $0.54\pm0.15$ & $10.61\pm0.93$ & $10.03\pm0.44$ & - \\
  NGC~4631 & 0.647 & $45.92\pm0.71$ & $<0.07$ & $10.15\pm0.13$ & $<0.27$ & $<1.47$ & $<3.67$ & $85.16\pm0.76$ & $114.74\pm1.89$ & - \\
  NGC~4666 & 0.517 & $35.65\pm2.01$ & $<0.57$ & $8.32\pm0.15$ & $<0.89$ & $6.77\pm1.94$ & $5.14\pm0.32$ & $68.92\pm1.84$ & $124.15\pm2.25$ & - \\
  NGC~4736 & 0.734 & $3.74\pm0.34$ & $<0.12$ & $4.25\pm0.07$ & $<0.76$ & $2.01\pm0.36$ & $3.22\pm0.44$ & $6.38\pm1.50$ & $20.28\pm1.13$ & - \\
  NGC~4818 & 0.314 & $195.95\pm14.93$ & $<0.38$ & $13.37\pm0.73$ & $<1.62$ & $<4.39$ & $8.86\pm0.50$ & $86.31\pm3.67$ & $125.45\pm3.74$ & - \\
  NGC~4945 & 0.712 & $698.41\pm60.43$ & $7.06\pm0.31$ & $68.07\pm2.27$ & $<2.12$ & $28.35\pm1.39$ & $49.26\pm13.00$ & $359.62\pm20.42$ & $732.76\pm7.84$ & Y \\
  NGC~5033 & 0.644 & $13.26\pm0.18$ & $0.42\pm0.05$ & $5.08\pm0.15$ & $0.48\pm0.08$ & $5.08\pm0.51$ & $2.02\pm0.47$ & $17.38\pm0.85$ & $45.35\pm1.52$ & Y \\
  NGC~5055 & 0.606 & $6.04\pm0.22$ & $<0.11$ & $3.08\pm0.10$ & $<0.89$ & $1.75\pm0.16$ & $2.19\pm0.47$ & $11.90\pm0.65$ & $34.10\pm1.84$ & - \\
  NGC~5128 & 0.165 & $202.71\pm3.97$ & $21.96\pm0.88$ & $149.98\pm5.04$ & $27.88\pm0.71$ & $123.63\pm2.72$ & $24.99\pm5.72$ & $140.69\pm6.05$ & $297.43\pm1.52$ & Y \\
  NGC~5194 & 0.591 & $17.03\pm0.26$ & $0.74\pm0.10$ & $10.91\pm0.24$ & $1.61\pm0.11$ & $7.93\pm0.25$ & $4.33\pm0.87$ & $18.39\pm0.81$ & $60.42\pm2.08$ & Y \\
  NGC~5195 & 0.842 & $5.54\pm0.72$ & $0.20\pm0.06$ & $2.17\pm0.11$ & $<0.39$ & $0.94\pm0.23$ & $<0.96$ & $3.22\pm0.39$ & $12.21\pm1.26$ & Y \\
  NGC~5236 & 0.602 & $503.33\pm19.88$ & $<0.61$ & $29.30\pm0.77$ & $<1.19$ & $5.75\pm1.08$ & $18.57\pm3.42$ & $263.50\pm9.21$ & $391.40\pm8.55$ & - \\
  NGC~5643 & 0.227 & $46.41\pm3.56$ & $24.63\pm1.03$ & $56.47\pm1.16$ & $31.33\pm1.77$ & $118.28\pm4.74$ & $<4.00$ & $33.26\pm3.05$ & $42.67\pm2.84$ & Y \\
  NGC~5907 & 0.791 & $6.07\pm0.22$ & $<0.01$ & $1.28\pm0.16$ & $<0.07$ & $1.57\pm0.51$ & $0.94\pm0.20$ & $14.21\pm1.41$ & $23.94\pm0.60$ & - \\
  NGC~6300 & 0.125 & $11.52\pm0.63$ & $12.54\pm0.66$ & $15.28\pm0.57$ & $8.33\pm1.29$ & $29.45\pm2.26$ & $<3.49$ & $<7.63$ & $11.31\pm0.82$ & Y \\
  NGC~6744 & 0.245 & $1.06\pm0.26$ & $<0.03$ & $1.53\pm0.17$ & $<0.13$ & $<0.73$ & $0.84\pm0.25$ & $3.57\pm0.98$ & $3.24\pm0.41$ & - \\
  NGC~6946 & 0.768 & $38.45\pm0.66$ & $<0.17$ & $3.77\pm0.10$ & $<0.59$ & $4.00\pm0.68$ & $8.42\pm0.60$ & $61.00\pm3.70$ & $112.25\pm1.57$ & - \\
  NGC~7331 & 0.337 & $4.47\pm0.15$ & $<0.16$ & $3.08\pm0.10$ & $<0.39$ & $1.88\pm0.14$ & $1.61\pm0.28$ & $13.81\pm0.41$ & $30.58\pm0.93$ & - \\
  UGCA~127 & 0.982 & $9.34\pm0.38$ & $<0.15$ & $1.53\pm0.24$ & $<0.97$ & $3.37\pm0.16$ & $3.43\pm0.12$ & $29.37\pm2.41$ & $44.08\pm1.68$ & - \\
\hline\end{tabular}
\end{center}
\medskip
{\tiny NOTES:} (1) Common galaxy name. (2) Equivalent width of the
$11.3\um$ PAH feature in units of $\um$. (3--10) Fluxes and their
statistical uncertainties for the measured mid-IR narrow emission
lines in units $10^{-14} \ergpcmsqps$. The mean uncertainty of the
fluxes is approximately 10 percent. $3\sigma$ upper limits are quoted
for non-detections. (11) Mid-IR AGN on the basis of \nev emission.
\end{minipage}
\end{table}
\end{landscape}


\noindent For the reduction of the IRS-spectral mapping data, custom
bad-pixel masks were again applied to the BCD images. {\small CUBISM}
\citep{cubism} was then used in conjunction with the latest flux
calibration files to construct final data cubes and extract spectra of
the central regions matched to the sizes of projected SH and LH
apertures.

After extraction of the raw spectra, the ends of each echelle order
were trimmed to remove the additional spectral noise caused by the
poor response of the grating.\footnote{Wavelength trim ranges are
given in Table 5.1 of the \spitzer-IRS Observers Manual.} Using the
redshifts given in the RBGS, each extracted spectrum was shifted to
rest-wavelength for further spectral analysis.

Solely for presentation purposes, single continuous spectra of each
slit were produced. Echelle orders were matched by fitting each
spectral continuum from a given order with either a first or second
order polynomial. Each echelle order continuum was then combined by
matching and calibrating to the 1st echelle order of the relevant slit
to construct the final SH and LH continua.  Due to the different
aperture sizes (and hence continuum fluxes), we have not attempted to
match the SH and LH spectra. Fig. \ref{fig_spectra01} shows the
reduced spectra for each of the sources using the SH and LH slits
(left and right panels respectively).

\subsection{Measuring Emission-line Properties}

Second-order polynomials were used to model the continuum and gaussian
profiles were simultaneously fit to spectral features to determine
emission-line fluxes. These were calculated using the IDL-based
spectral analysis tool {\small SMART} \citep{smart}.\footnote{SMART
was developed by the IRS Team at Cornell University and is available
through the {\it Spitzer} Science Center at Caltech.} Fluxes or $3
\sigma$ upper-limits are given in Table 3 for the following emission
lines: [NeII] $\lambda 12.81\um$, [NeV] $\lambda 14.32\um$, [NeIII]
$\lambda 15.56\um$, [NeV] $\lambda 24.32\um$, [OIV] $\lambda
25.89\um$, [FeII] $\lambda 25.99\um$, [SIII] $\lambda 33.48\um$ and
[SiII] $\lambda 34.82\um$). Polycyclic Aromatic Hydrocarbon (PAH;
e.g.,\ \citealt{draine03}) features are detected in many of the
galaxies; we use the $11.3$~$\um$ PAH equivalent width in our analyses
and report these values in Table~3. Due to their broad profiles, the
strength of the PAH features are measured using multiple
gaussians. The results obtained from our continuum and emission-line
fitting procedure are shown in Fig. \ref{fig_nevfits} for all \nev
detected galaxies.

Of the 64 galaxies in the $D < 15$~Mpc sample, 16 have published
\spitzer-IRS data in S08 and 18 have published \spitzer-IRS data in
\citeauthor{dale09} (2009; hereafter D09). However, due to differing
data-reduction routines and approaches in the detection of emission
lines (e.g.,\ we measured emission-line properties in apertures
centred on the near-IR nucleus, S08 searched for emission lines in
small apertures across the circumnuclear region of each galaxy, and
D09 constrained emission-line properties in large apertures across the
extent of each galaxy), we have re-analysed all of the galaxies to
provide self-consistent results. However, despite these differing
approaches, we find average emission-line flux variances of only
$\approx$~10--30\% between our study and that of S08 and D09 for those
galaxies with with detected [NeII] $\lambda 12.81\um$, [NeIII]
$\lambda 15.56\um$, and [OIV] $\lambda 25.89\um$ emission; when
comparing to D09 we scaled our emission-line fluxes by the difference
in aperture size between D09 and our study.


\section{Results}

\begin{figure*}
\vspace{-1.3cm}
\begin{center}
\hspace{0.0cm}
\vspace{-1.3cm}
\includegraphics[width=0.85\textwidth]{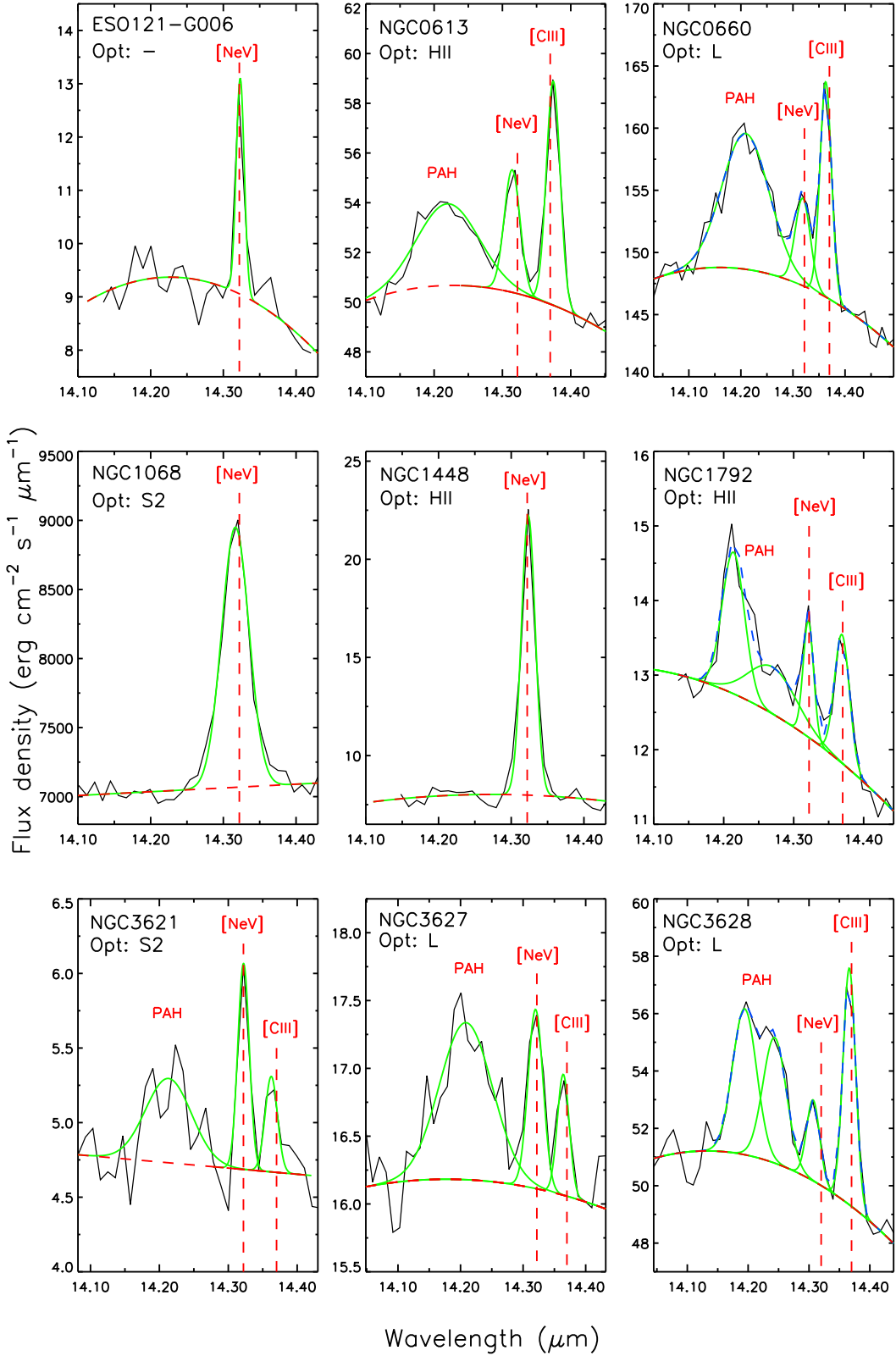}
\vspace{0.8cm}
\caption{Mid-IR spectra expanded around the \nev emission-line region
for all [NeV] detected galaxies covering a broad range of optical and
mid-IR spectral properties. The dotted line indicates the second-order
polynomial fit to the continuum and the solid lines indicate the
gaussian fit to the emission-line features; the \nev, $\lambda 14.2
\um$ PAH, and [ClII] $\lambda 14.37 \um$ emission features are labeled
when detected. The identification of AGN activity from the \nev
emission-line in ten objects is complicated by the PAH and [CIII]
star-forming signatures and thus high signal-to-noise data is
required.}
\label{fig_nevfits}
\end{center}
\end{figure*}

\begin{figure*}
\vspace{-1.3cm}
\begin{center}
\hspace{1.2cm}
\vspace{-1.3cm}
\includegraphics[width=0.85\textwidth]{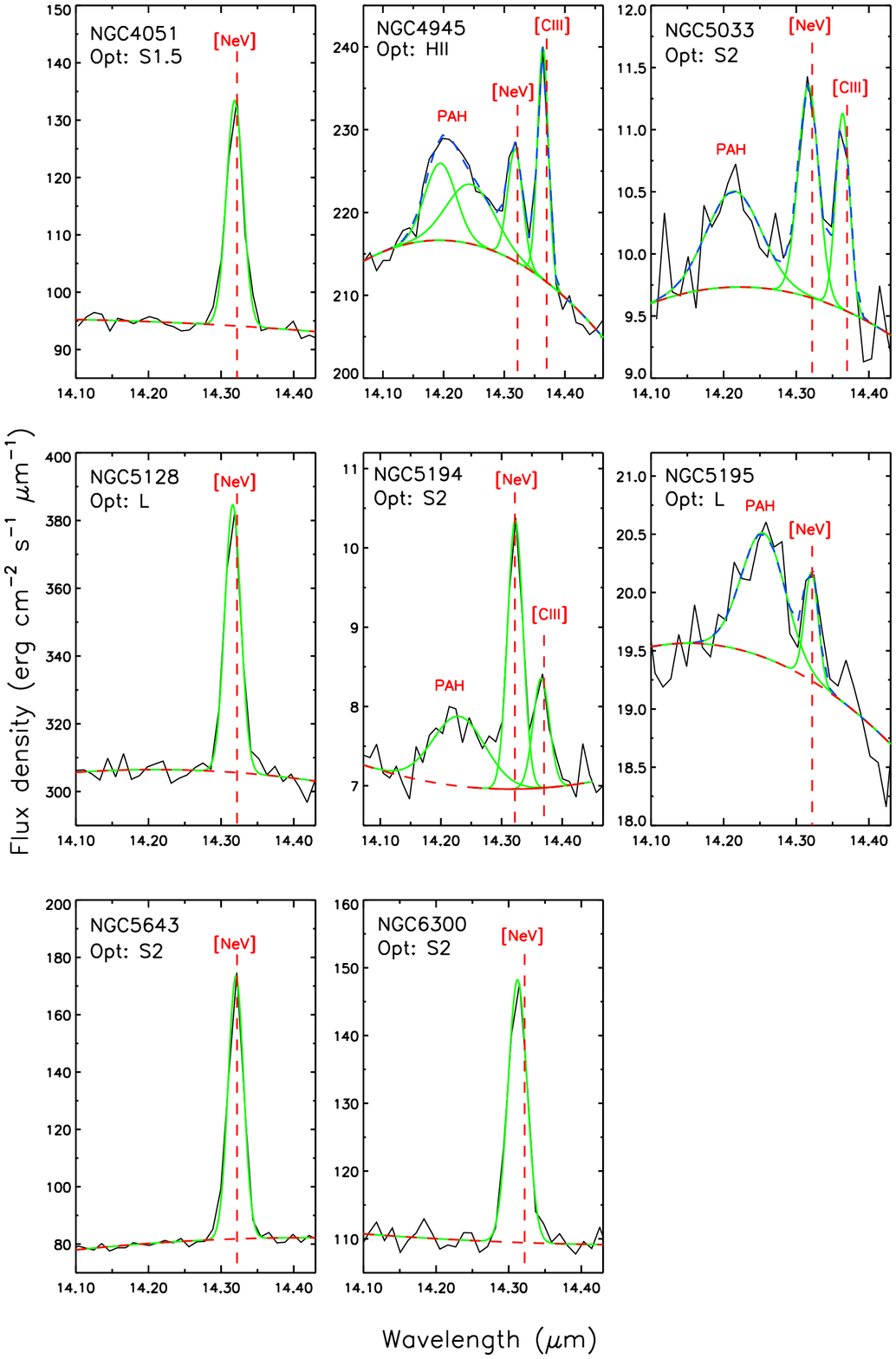}
\vspace{0.8cm}
\contcaption{}
\end{center}
\end{figure*}

\subsection{The Discovery of a Significant Population of Optically Unidentified AGNs}

\begin{figure*}
\begin{center}
\hspace{-1.0cm}
\vspace{-1.0cm}
\includegraphics[width=\textwidth]{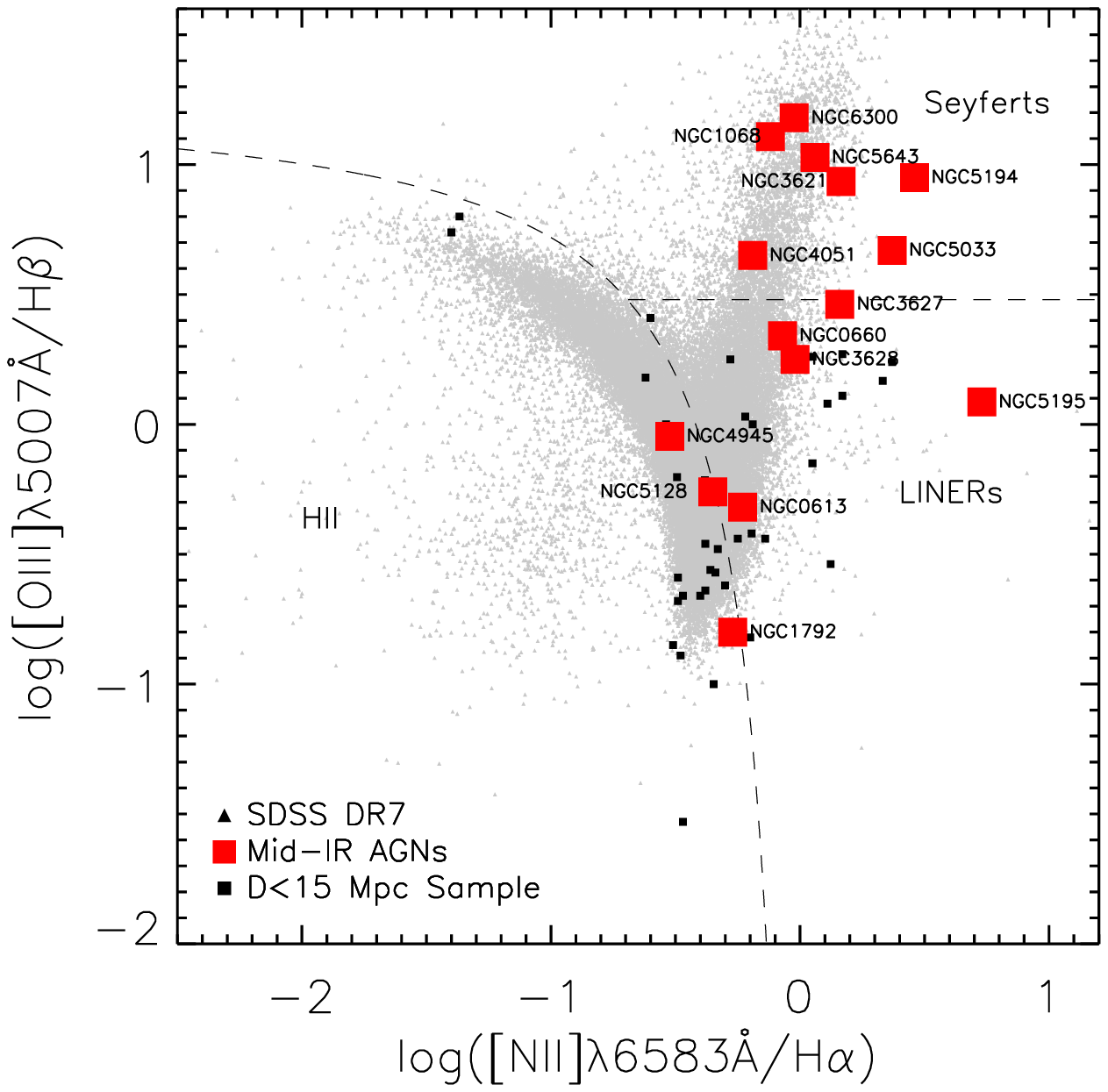}
\vspace{0.5cm}
\caption{Emission-line diagnostic diagram showing ${\rm [OIII]}\lambda
5007 {\rm \AA} /{\rm H}\beta$ versus ${\rm [NII]}\lambda 6583 {\rm
\AA} /{\rm H}\alpha$ for our $D<15$ Mpc galaxy sample (squares) and
the Seventh Data Release of SDSS (dots). Kauffmann et al. (2003)
demarcation lines are shown to isolate the Seyfert, LINER and
star-forming HII regions (dashed lines). Those galaxies with $3\sigma$
\nev detections (i.e.,\ mid-IR identified AGNs) are labeled and
highlighted with large squares. Seven of the $D<15$ Mpc galaxies are
optically classified as Seyferts, all of which are identified as
mid-IR AGNs, while the other objects are optically classified either
as HII galaxies or LINERs. Two additional mid-IR AGNs (NGC~1448;
ESO121-G006) are not shown in the figure. NGC~1448 is classified as an
HII galaxy (Veron-Cetty et~al. 1986) but cannot be plotted on this
figure since both [OIII]$\lambda5007$ and H$\beta$ are
undetected. ESO121-G006 lacks good-quality optical spectroscopy but
since it is hosted in a highly inclined galaxy, it is likely to be an
optically unidentified AGN; see \S3.3 and Fig.~\ref{fig_agn_images}.}
\label{fig_bpt}
\end{center}
\end{figure*}

\begin{figure*}
\begin{center}
\includegraphics[width=8.5cm]{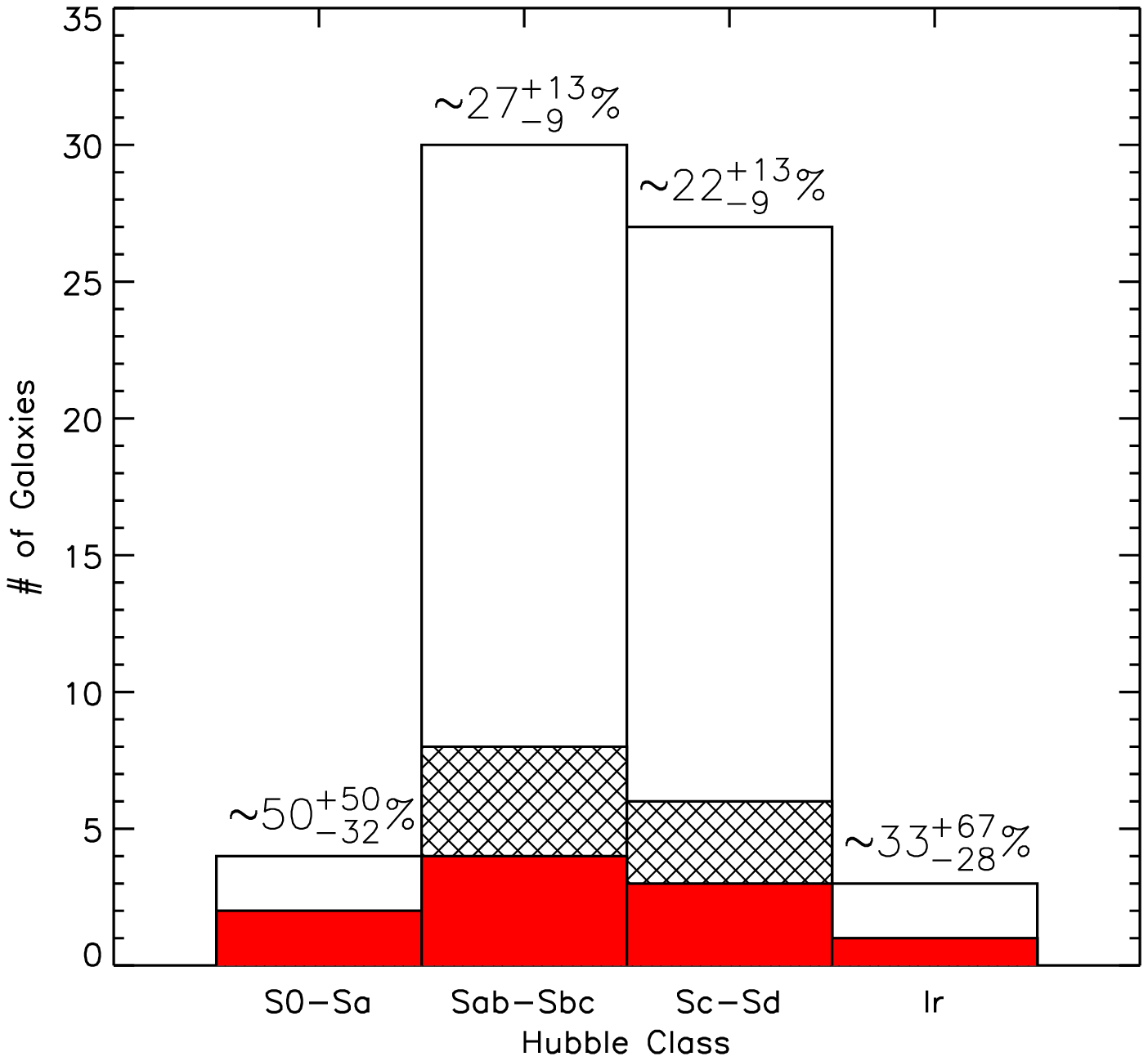}
\includegraphics[width=8.5cm]{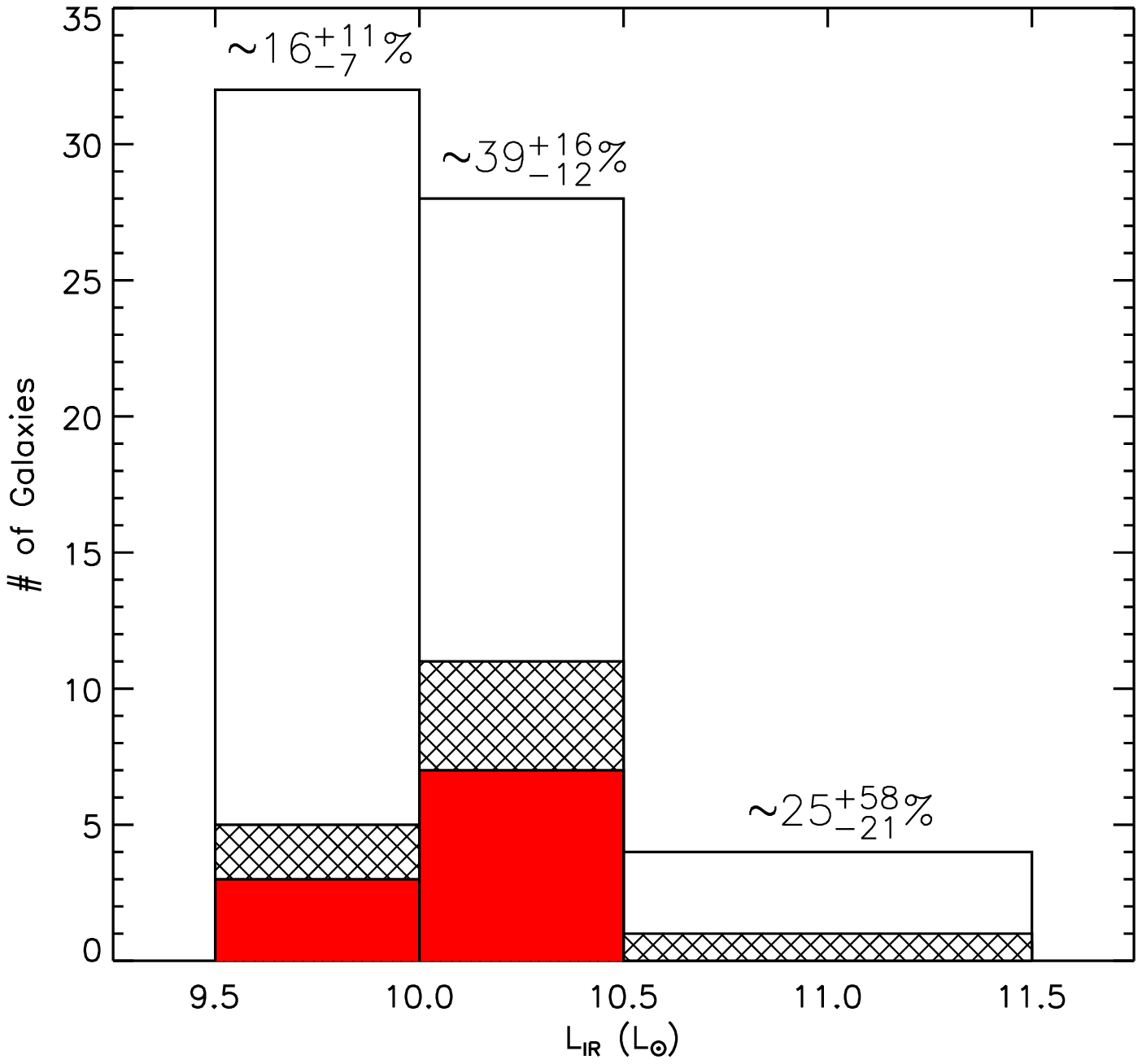}

\caption{Fraction of AGNs identified in the $D<15$ Mpc sample shown as
a function of host-galaxy morphological classification and IR
luminosity; the AGN fraction and associated $1 \sigma$ errors are
given for each sample bin. We further sub-divide the AGNs into \nev
detected sources which are also optically observed to have AGN
signatures (cross hatching), \nev detected sources which lack optical
AGN signatures (solid colour), and combining these to give all
galaxies with detected \nev emission, i.e. total number of AGN
(cross-hatch+solid colour).}
\label{morph_mbh}
\end{center}
\end{figure*}

Of the 64 objects in our $D<15$~Mpc sample (presented in Tables 1 and
3), 17 have $3\sigma$ detections of the [NeV] $\lambda 14.32 \um$
emission line, and therefore have unambiguous evidence for AGN
activity.\footnote{Theoretical modeling has shown that [NeV] can also
  be produced in galaxies containing a large population of Wolf-Rayet
  (WR) stars \citep{schaerer99}. Two WR galaxies are present in our
  volume-limited sample (IIZw40 and NGC~1569). However, [NeV] emission
  is not detected in either source, and their mid-IR continuum and
  spectral features are clearly distinct from those of AGNs.} We
therefore find an overall AGN fraction of $\approx 27^{+8}_{-6}$
percent in the most bolometrically luminous galaxies to $D< 15$ Mpc
($\Lir\approx(0.3-20)\times10^{10} \Lsun$).\footnote{Small-number
  Poisson statistical errors are calculated for upper and lower limits
  based on the $1 \sigma$ confidence levels given in
  \citet{gehrels86}.}

Of the 16 galaxies in common between S08 and our sample, S08
identified [NeV] (14.32 or 24.32~$\um$) in four systems (NGC~3556;
NGC~3938; NGC~4536; NGC~5055). We do not detect significant [NeV] in
any of these four galaxies, which may be due to differences in the
positions and sizes of the apertures used to extract the spectra
within {\small CUBISM}. Of the 18 galaxies in common between D09 and
our sample, D09 identified [NeV] $\lambda 14.32 \um$ in three systems
(NGC~3621; NGC~5033; NGC~5194), all of which we also identify here;
however, we also identified [NeV] emission in two systems where D09
quote [NeV] upper limits (NGC~3627; NGC~5195), which could be due to
dilution of the [NeV] emission by the host galaxy in the large
apertures used by D09. We note that differences in the identification
of weak [NeV] emission can also be due to the adopted emission-line
detection procedure and signal-to-noise ratio threshold.

Seven ($\approx 11^{+6}_{-4}$ percent) of the 64 galaxies in our
sample are unambiguously identified as AGNs using classical optical
emission-line diagnostics (i.e.,\ optically classified as Seyfert
galaxies; e.g.,\ \citealt{bpt}; \citealt{veil_oster87};
\citealt{kauff03a}); see Table~1. All of these optically identified
AGNs are classified as AGNs at mid-IR wavelengths in our analysis. In
Fig. \ref{fig_bpt}, we plot the optical emission-line ratios of the 53
galaxies with good-quality optical spectra; of the other 11 galaxies
in the sample, three do not have a sufficient number of detected
emission lines to classify at optical wavelengths using emission-line
diagnostics, and eight do not have published optical spectroscopy. Ten
of our mid-IR classified AGNs are not unambiguously identified as AGNs
at optical wavelengths: five are classified as HII galaxies (NGC~0613,
NGC~1448, NGC~1792, NGC~4945, and NGC~5128), four are classified as
LINERS (NGC~0660, NGC~3627, NGC~3628, and NGC~5195), and one does not
have good-quality optical spectroscopic data (ESO121-G006). Although
ESO121-G006 does not have a good-quality optical spectrum, since it is
hosted in a highly inclined galaxy we would expect it to be an
optically unidentified AGN; see \S3.3 and
Fig.~\ref{fig_agn_images}. Large optical spectroscopic studies (e.g.,
Veilleux et~al. 1995; Ho97) have speculated that the central regions
of many LINERs are likely to be powered by AGN activity. Indeed, using
{\it Chandra} X-ray observations, Ho et~al. (2001) suggest $\approx
60$ percent of all LINERs host AGNs. By contrast, using our mid-IR
diagnostics, we find only $\approx 25$ percent of IR-bright LINERs
appear to be AGNs, which could be due to a number of factors (e.g.,\
different sample selection and ambiguous evidence for AGN activity at
X-ray energies; i.e.,\ X-ray binaries).

The optical spectroscopy for the galaxies in our sample comes from a
variety of different studies with a range in emission-line
sensitivities. However, the majority of the galaxies are in Ho97,
which arguably comprises the most sensitive optical spectroscopy of a
large number of nearby galaxies. Of the 38 galaxies present in both
our sample and Ho97, we find that eight are mid-IR identified AGNs
(NGC~0660, NGC~1068, NGC~3627, NGC~3628, NGC~4051, NGC~5033, NGC~5194
and NGC~5195), only four of which are unambiguously identified as AGN
at optical wavelengths. One of the optically unidentified AGNs is
NGC~5195, which may be a binary AGN system with NGC~5194 in the
Whirlpool galaxy. Potential AGN activity has also been found in
NGC~5195 using \chandra \ observations \citep{terashima04}.

Optical spectroscopic surveys have found that AGNs typically reside in
moderately massive bulge-dominated galaxies (Hubble-type: E--Sbc;
$M_{\star} \approx$~(0.1--3)~$\times 10^{11} \Msun$; e.g.,\ Ho97;
\citealt{heck04}). In Fig. \ref{morph_mbh} we show the histogram of
galaxy morphology for our $D<15$ Mpc sample. We find that the host
galaxies of our mid-IR identified AGNs cover a wide range of galaxy
type (S0--Ir). However, in contrast to Ho97 and \citet{heck04}, we
find that a large fraction of Sc--Sd-type galaxies host AGN activity
at mid-IR wavelengths ($\approx 22^{+13}_{-9}$ percent; i.e.,\ a
comparable AGN fraction to that found in Sab--Sbc galaxies). This
shows that late-type galaxies typically assumed to host pseudo bulges
(Sc--Sd) can harbour AGN activity, and therefore must host a SMBH. As
found in previous studies, this indicates that galaxies without
classical bulges can host SMBHs (e.g.,\ \citealt{greene08}; Barth
et~al. 2009).

\begin{figure*}
\begin{center}
\hspace{-1.0cm}
\includegraphics[width=9.5cm]{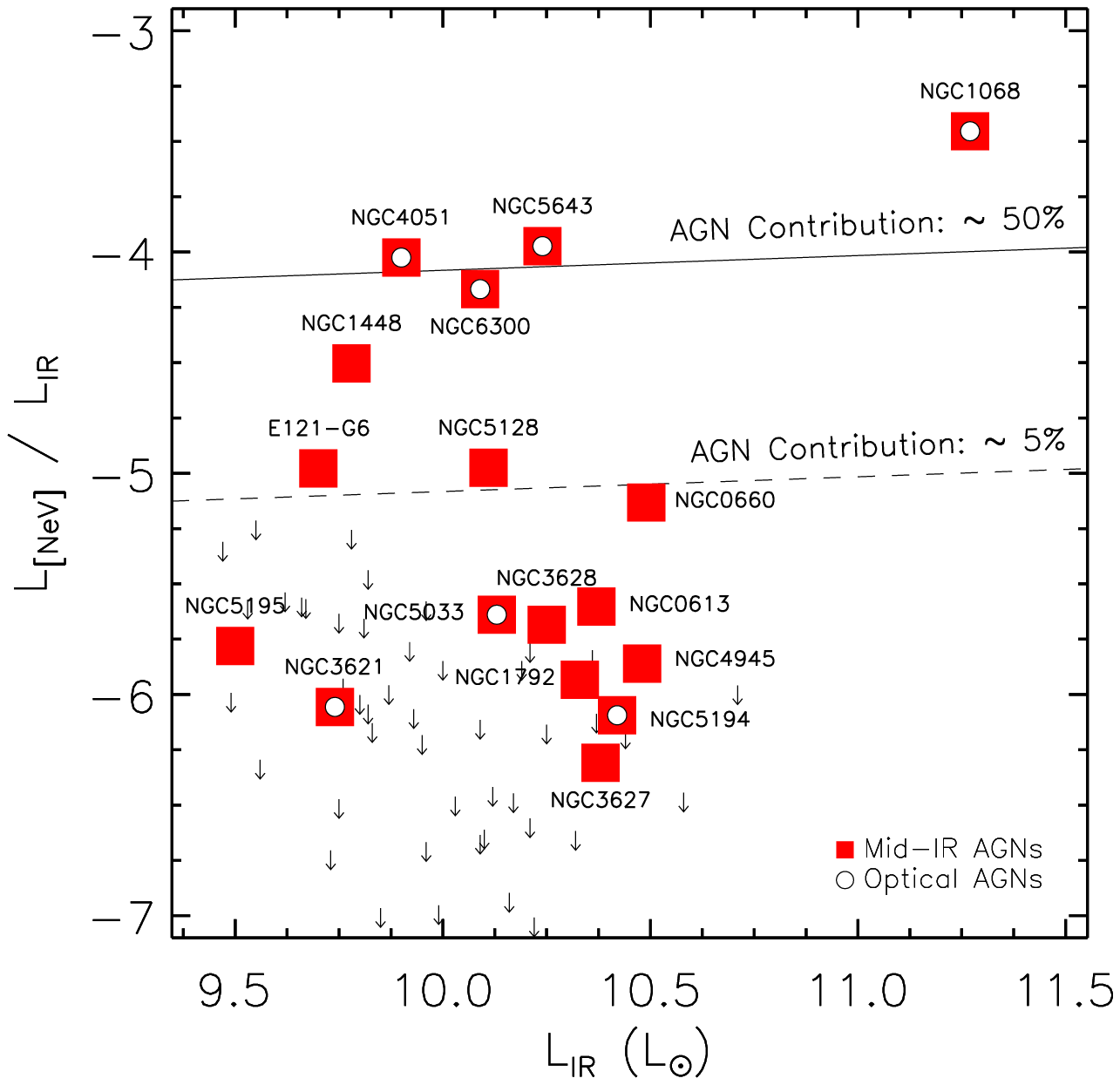}
\hspace{-1.0cm}
\includegraphics[width=9.5cm]{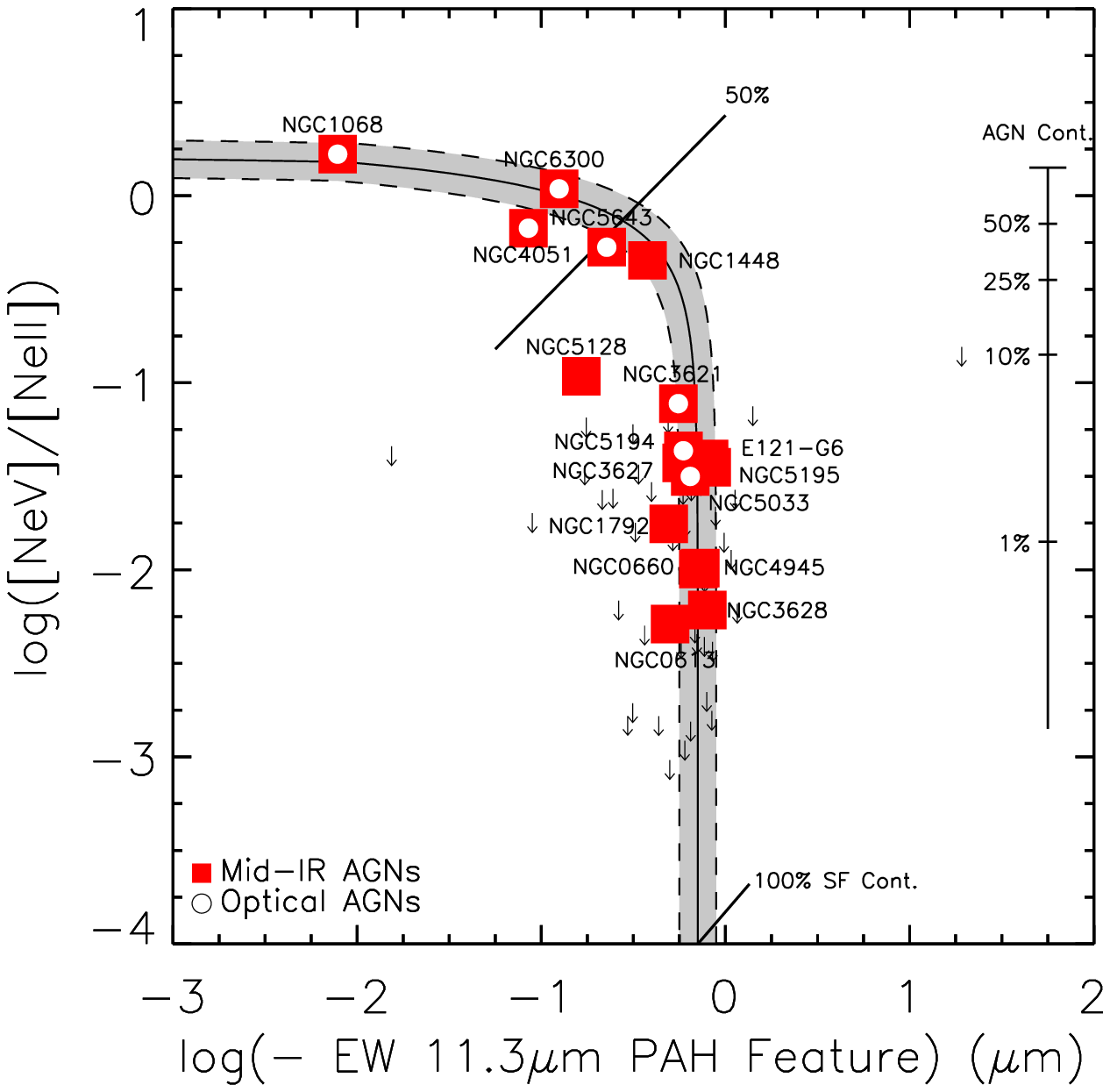}
\caption{{\bf a; left):} \nev--IR luminosity ratio versus IR
luminosity. Mid-IR identified AGNs (filled squares) and optically
identified AGNs (open circles) are indicated. The solid and dashed
lines show the expected luminosity ratio for a $\approx$~50 percent
and $\approx$~5 percent AGN contribution to the IR emission,
respectively; adapted from \citet{sat07}. Both the optically
identified and unidentified AGNs span a wide range of [NeV]--IR
luminosity ratios. However, clearly the {\it Spitzer}-IRS observations
are comparatively less sensitive towards the detection of \nev
emission in the lowest IR luminosity bin ($\Lir<10^{10}$~$\Lsun$),
indicating that the overall AGN fraction given in Fig.~5 should be
considered a lower limit. {\bf b; right):} Flux ratio of the \nev and [NeII]
$\lambda 12.82 \um$ emission lines versus equivalent width of the
$11.3 \um$ PAH feature (EW$_{\rm 11.3,PAH}$). Both the EW$_{\rm
11.3,PAH}$ and the [NeV]--[NeII] flux ratio give relative measures of
the AGN--starburst contribution. The vertical scale indicates the
fraction of AGN activity for the [NeV]--[NeII] flux ratio
\citep{sturm02}. A linear mixing model curve (solid line) and 0.1 dex
errors (dashed) are shown: a 100\% AGN is assumed to have [NeV]/[NeII]
$\sim 1.4$, EW$_{\rm 11.3,PAH}$$\sim 0.008$~$\mu$m, while a 100\%
starburst is assumed to have [NeV]/[NeII] $\sim 0.001$, EW$_{\rm
11.3,PAH}$$\sim 0.7$~$\mu$m.}
\label{fig_mir_bpt}
\label{fig_lir_nev}
\end{center}
\end{figure*}

In Fig. \ref{morph_mbh} we also show the incidence of AGN activity as
a function of IR luminosity. In the moderate-luminosity IR bin ($\Lir
\approx$~(1--3)~$\times10^{10} \Lsun$), we find a large AGN fraction
of $\approx 39^{+16}_{-12}$ percent, suggesting that the overall AGN
fraction for our sample may be a lower limit. The smaller AGN
fractions found in the lower ($\Lir < 10^{10} \Lsun$; $\approx
16^{+11}_{-7}$ percent) and higher ($\Lir >3\times10^{10} \Lsun$;
$\approx 25^{+58}_{-21}$ percent) IR luminosity bins could be due to
relatively weaker AGN sensitivity limits (i.e.,\ a higher ${\rm L_{\rm
[NeV]}}$/$\Lir$ emission-line ratio) and large uncertainties due to
small-number statistics, respectively. Indeed, we find similar AGN
fractions (of order $\approx$~10 percent) in both the lower IR
luminosity and moderate IR luminosity bins if we only consider AGNs
identified with ${\rm L_{\rm [NeV]}}$/$\Lir$$<10^{-5}$, suggesting
that further AGNs remain to be detected in the lower IR luminosity;
see Fig.~\ref{fig_lir_nev}a. This may indicate that the overall AGN
fraction in our IR-bright sample may be closer to $\approx$~40
percent. 

The large AGN fraction found in our study indicates a tighter
connection between AGN activity and IR luminosity for galaxies in the
local Universe than previously found, exceeding the AGN fraction
obtained with optical spectroscopy by up-to an order of magnitude
(e.g.,\ compared to the results for $\Lir <10^{11} \Lsun$ galaxies in
\citealt{veilleux99}). This may indicate a close association between
AGN activity and star formation, as is typically expected given the
tight relationship between SMBH and spheroid mass in the local
Universe (e.g.,\ Magorrian et~al. 1998; Gebhardt et~al. 2000). There
are probably two reasons why we identify a significantly larger AGN
fraction than previously found: (1) mid-IR spectroscopy provides a
more sensitive probe of AGN activity than optical spectroscopy, and
(2) our galaxies are very nearby, allowing us to identify faint AGN
signatures.

\subsection{Why are AGN signatures often absent at optical wavelengths?}

Of the seventeen galaxies in our $D<15$~Mpc sample that unambiguously
host AGN activity, ten ($\approx 60^{+25}_{-18}$ percent) lack AGN
signatures at optical wavelengths. Here we explore the three most
likely reasons why the AGN signatures are absent in the optical
spectra of these galaxies: (1) the optically unidentified AGNs are
intrinsically lower luminosity systems, (2) the optically unidentified
AGNs have a larger fraction of star formation/stellar light that
dilutes the optical AGN signatures, or (3) the optically unidentified
AGNs are more heavily obscured at optical wavelengths.

\subsubsection{Are optically unidentified AGNs intrinsically low luminosity?}

The non detection of optical AGN signatures in the optically
unidentified AGNs may be due to the AGNs being lower luminosity
systems. In Fig. \ref{fig_oiii_nev}a, we show the \nev luminosities of
the mid-IR identified AGNs and a sample of well-studied local AGNs
from \citet{panessa06} as a function of their [OIII] $\lambda
5007$\AA \ luminosity. We characterise the [NeV]--[OIII] luminosity
relationship using a regressional fit.\footnote{The fit to the data
was obtained using the IDL-based robust bi-sector linefit algorithm
{\small ROBUST\_LINEFIT}.}

\begin{equation}
{\rm log [NeV]} = (1.19 \pm 0.12) \ {\rm log [OIII]} - (8.65 \pm 4.60)
\end{equation}

The tightness in the [NeV]--[OIII] luminosity relationship indicates
that the [NeV] luminosity provides a reliable measurement of the
intrinsic luminosity of the AGN; the optically identified AGNs lie on
this relationship if we correct the [OIII] luminosity for extinction
as measured using the Balmer decrement (see \S3.2.3).

The optically identified AGNs in our sample cover a broader range of
\nev luminosities than the optically unidentified AGNs ($\Lnev \approx
10^{37}$--$10^{41} \ergps$ and $\Lnev \approx 10^{37}$--$10^{39}
\ergps$, respectively). However, since there are optically identified
AGNs with similar luminosities to the optically unidentified AGNs,
this indicates that the dominant reason for the non detection of the
optically unidentified AGNs cannot be due to them hosting
intrinsically lower-luminosity AGN activity. Fig. \ref{fig_oiii_nev}a
also shows that the non identification of optical AGN signatures in
the optically unidentified AGNs is not due to low-sensitivity optical
spectroscopy since the [OIII]/[NeV] luminosity ratios are lower than
that given in Equation 1.

\subsubsection{Are optically unidentified AGNs star-formation dominated?}
\label{sec:starform}

The non-identification of AGNs at optical wavelengths could be due to
dilution from star-formation signatures, which we can test with our
data. The mid-IR continua of starburst galaxies are typically
characterised by strong PAH features at $\lambda \sim 3.3$, 6.2, 7.7,
8.6, 11.3, 12.7, 14.2 and $17.0 \um$ combined with a steep spectral
slope at far-IR ($\lambda > 25 \um$) wavelengths \citep{brandl06}. By
contrast, PAH features tend to be weak or absent in AGN-dominated
systems (e.g., NGC~1068; \citealt{rigopoulou02}). It is apparent from
the \spitzer-IRS spectroscopy presented in Fig.~\ref{fig_nevfits} and
\ref{fig_spectra01}, that most of the optically unidentified AGNs
exhibit star-formation signatures at mid-IR wavelengths, indicating
that they host joint AGN--starburst activity. However, to test whether
the lack of AGN signatures at optical wavelengths is due to dilution
from star-formation, we need to compare the relative
AGN--star-formation contributions for both the optically identified
and unidentified AGNs.

As the sources in the sample are well resolved, we can use the {\it
Spitzer}-IRS spectra to quantify the relative strengths of the
star-formation and AGN activity in the circumnuclear regions of the
mid-IR identified AGNs. Previous studies have shown that the
equivalent width of the IR-detected PAH features at 6.2 and $7.7\um$
are well correlated with the AGN-starburst activity occurring within a
galaxy (e.g.,\ \citealt{genzel98}; \citealt{laurent00};
\citealt{peeters04}; \citealt{dale06}). However, given the spectral
coverage of the \spitzer-IRS spectroscopy for the $D<15$~Mpc sample
($\lambda\sim$~9.9--37.2~$\mu$m), here we calibrate and use the
equivalent width of the $11.3 \um$ PAH feature (EW$_{\rm 11.3,PAH}$,
which is detected in every galaxy) to indicate the relative
AGN--star-formation contribution of each galaxy. \citet{martin06} find
that for a starburst-dominated circumnuclear region EW$_{\rm
11.3,PAH}$ (100 percent SF) $\approx 0.7$~$\mu$m. Comparing this to an
AGN-dominated system (i.e.,\ NGC~1068) we find EW$_{\rm 11.3,PAH}$
(100 percent AGN) $\approx 0.008$~$\mu$m. The mid-IR emission line
ratios can also be used to constrain the relative contributions of AGN
and stellar emission (e.g.,\ Dale et al. 2006). Since [NeII] $\lambda
12.82 \um$ emission is primarily produced by star-formation activity
within a galaxy, and \nev is solely attributed to AGN activity, the
ratio of these emission lines is also a strong tracer of the relative
strengths of these two processes (e.g., Sturm et al. 2002).

In Fig.~\ref{fig_mir_bpt}b we present the predicted EW$_{\rm
  11.3,PAH}$ and [NeII] $\lambda 12.82 \um$ -- \nev \ flux ratios for
different AGN--star-formation contributions; this is similar in
principal to Genzel et~al. (1998). Both of the AGN--star-formation
ratio estimates are in good agremment for the majority of the mid-IR
AGNs, with a mean dispersion of $\approx 20$ percent. We note that the
predicted AGN--star-formation contributions for four ($\approx 25$
percent) of the galaxies (NGC~1068, NGC~4051, NGC~5128 and NGC~5643)
differ by more than a factor of two from the mixing model, whilst the
other mid-IR identified AGNs lie within a factor of 40 percent. One of
these such outliers is the FR-1 radio galaxy, NGC~5128 (EW$_{\rm
  11.3,PAH,model} > 4\times$EW$_{\rm 11.3,PAH,obs}$), which could be
due to dilution of the PAH feature by an underlying synchrotron
component (related to the radio emission) that is emitting at mid-IR
wavelengths. On the basis of Fig.~\ref{fig_mir_bpt}b, the IR emission
for five ($29^{+20}_{-13}$ percent) of the mid-IR identified AGNs
(NGC~1068, NGC~6300, NGC~4051, NGC~5643 and NGC~1448) has a
significant contribution from AGN activity ($> 25$ percent), only one
of which is an optically unidentified AGN (NGC~1448). As may be
expected, these five galaxies also host the most luminous AGNs, as
shown in Fig. \ref{fig_oiii_nev}a. The IR emission for the other
twelve ($\approx71^{+26}_{-20}$ percent) mid-IR identified AGNs
appears to be star-formation dominated (AGN contribution $< 25$
percent), nine of which are optically unidentified AGNs. These
analyses have been performed using the SH module of {\it Spitzer}-IRS,
which traces only the circumnuclear region of these galaxies. However,
we get qualitatively similar results if we consider the $L_{\rm
  [NeV]}$/$\Lir$ ratios, which should provide a measure of the
contribution of the AGN to the {\it total} IR luminosity of the galaxy
(i.e., as measured by {\it IRAS}); see Fig.~\ref{fig_lir_nev}a.

On the basis of these analyses we therefore derive qualitatively
similar conclusions to those in \S3.2.1. Clearly, there is a
difference in the distribution of relative AGN--star formation
strengths, with the optically unidentified AGNs being typically more
star-formation dominated than the optically identified AGNs. However,
since approximately half of the optically identified AGNs are also
star-formation dominated, dilution from star-formation signatures is
unlikely to be the dominant cause for the lack of optical AGN
signatures in all of the optically unidentified AGNs. Indeed, as we
show in \S3.2.3 and \S3.3, strong H$\beta$ emission produced by young
stars is likely to be the primary reason for the lack of AGN optical
signatures (i.e., a low [OIII]--H$\beta$ ratio) in only three of the
optically unidentified AGNs.

\subsubsection{Are optically unidentified AGNs heavily dust obscured?}

\begin{figure*}
\begin{center}
\hspace{-1.0cm}
\includegraphics[width=9.5cm]{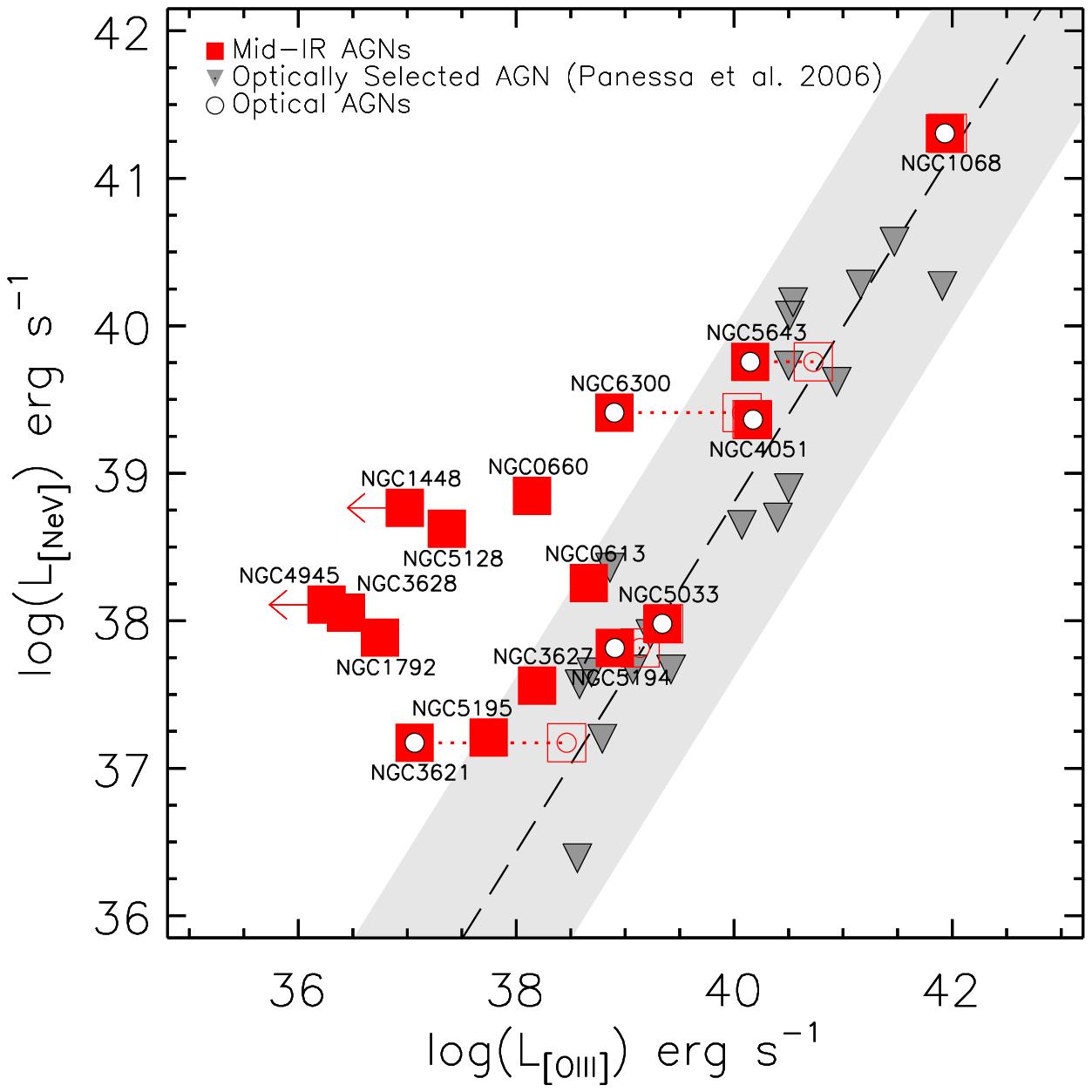}
\hspace{-1.0cm}
\includegraphics[width=9.5cm]{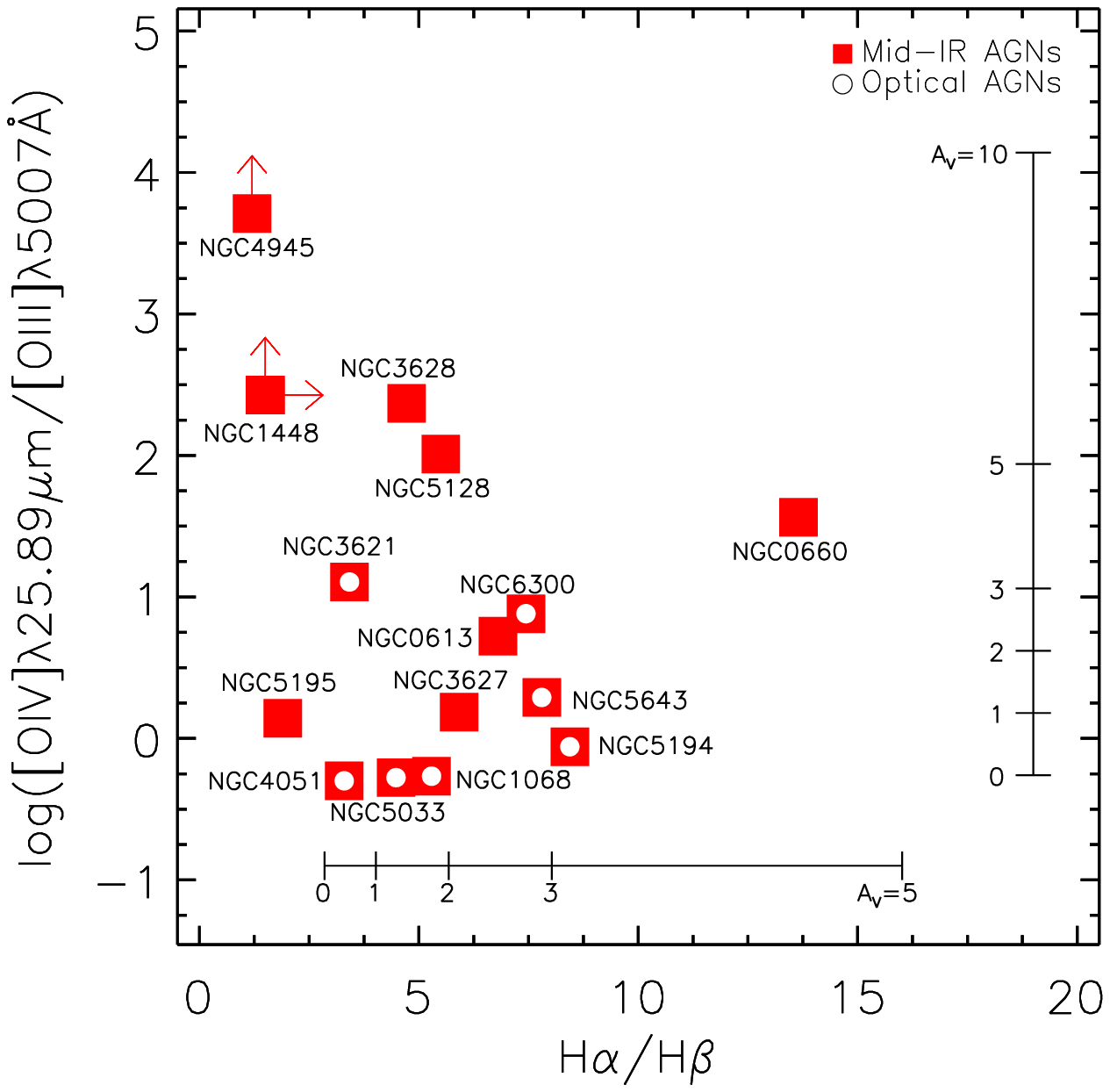}
\caption{{\bf a; left)} \nev luminosity versus [OIII] $\lambda
5007$\AA \ luminosity. Mid-IR identified AGNs (filled squares),
optically identified AGNs (open circles), and optically selected AGNs
from Panessa et~al. (2006; filled triangles) are indicated. The [OIII]
luminosities for the Panessa et~al. (2006) AGNs are corrected for
extinction, and the {\it Spitzer}-IRS data is reduced and analysed
following \S2. The dashed line was obtained using a robust bi-sector
linefit algorithm for the Panessa et~al. (2006) sample (see Equation
1) and shows that the \nev luminosity provides a reliable measurement
of the intrinsic luminosity of the AGN. The shaded region indicates
the intrinsic scatter within the relation. The [NeV]--[OIII]
luminosity ratios of the optically identified AGNs in our $D<15$ Mpc
sample are in good agreement with the [NeV]--[OIII] luminosity
relationship when the [OIII] luminosities are corrected for extinction
using the Balmer decrement (dotted lines; see \S3.2.3). Three of the
optically unidentified AGNs (NGC~613, NGC~3627 and NGC~5195) are
within the scatter of the relation indicating that these galaxies may
not necessarily be extinguished in [OIII] flux, but could have
enhanced H$\beta$ flux from young stars as suggested in
Fig~\ref{fig_lir_nev}b and \ref{fig_oiii_nev}b. {\bf b; right)} \oiv
and [OIII] $\lambda 5007$\AA \ flux ratio versus Balmer decrement
(${\rm H}{\alpha}/{\rm H}{\beta}$ flux ratio). These ratios are
indicators of the relative dust/gas extinction within the host
galaxies, although the [OIV]--[OIII] flux ratio should be more
sensitive towards high levels of extinction (see \S3.2.3 and
Diamond-Stanic et~al. 2009); the solid lines indicate the relationship
between the flux ratios and extinction in $A_{\rm V}$. The symbols
have the same meaning as in Fig.~6.}
\label{fig_oiii_nev}
\label{fig_extinc}
\end{center}
\end{figure*}

The dominant reason for the lack of AGN optical signatures could be
due to dust obscuration. A good measure of dust obscuration within a
host galaxy is the so-called Balmer decrement (the ${\rm H}
\alpha/{\rm H}\beta$ flux ratio; e.g., \citealt{ward87}). In Table 1
we show that both the optically identified and optically unidentified
AGNs cover similar ranges in Balmer decrements, apparently indicating
no difference in optical extinction between the two
populations. However, many late-type galaxies are found to be very
dust/gas rich, and for galaxies such as these, the Balmer decrement
may be a poor measure for high-levels of extinction. For example,
NGC~4945 is an optically unidentified AGN (optically classified as an
HII galaxy), and using analyses at near-IR wavelengths it is estimated
to have a V-band extinction of $A_V \approx 6$--20 magnitudes
\citep{moorwood96}; however, the optical Balmer decrement of the
galaxy would suggest that it is relatively unobscured ($A_V \approx 0$
mags). To remove this ambiguity of additional attenuation to the
optical emission, we require a measure of dust-obscuration that is
capable of probing greater optical depths than optical spectroscopy
alone.

Here we estimate the extinction towards the AGNs using \oiv at mid-IR
wavelengths and [OIII] $\lambda 5007$\AA\ at optical
wavelengths.\footnote{We note that it would also be useful to
determine the amount of extinction using the [NeV] emission lines at
$14.32 \um$ and $3426$\AA. Unfortunately, this is not currently
possible due to the lack of available data at ultra-violet wavelengths
for the majority of our sample. Also, this ratio is additionally
dependent on temperature variations of the photoionised gas within the
narrow line region of the AGN.} \oiv has often been used as a
relatively dust-obscuration independent indicator of the AGN
luminosity (e.g.,\ Genzel et~al. 1998; D06), and we show in \S3.4 that
it is reliable for [NeV]--identified AGNs; however, see \S3.4 for
caveats in using \oiv to directly identify AGN activity. By contrast,
the optically detected [OIII] $\lambda 5007$\AA\ emission lines
produced in the narrow-line regions of AGNs can be subject to strong
reddening by dust/gas, and therefore the [OIV]--[OIII] emission-line
ratio should provide a good indicator for the presence of heavy
obscuration. A similar approach using the [OIV]--[OIII] ratio has been
taken by Diamond-Stanic et~al. (2009) for assessing the obscuration in
a sample of X-ray luminous AGNs. There could also be some dependence
of the [OIV]--[OIII] ratio on the hardness of the radiation field
(e.g.,\ as measured using the [NeIII]--[NeII] ratio; Brandl
et~al. 2006); however, we find no strong dependence in our sample. We
calibrate the extinction correction factor in $A_V$ magnitudes by
assuming $f_{\rm [OIII]}$~${\propto}$~$f_{\rm [OIV]}$$^{1.8}$, as
found for Seyfert 2 galaxies (e.g.,\ \citealt{melendez08b}), combined
with the expected dust-reddening at $\lambda \approx 5007$\AA \ and
$25.9\um$ (e.g., \citealt{oster_book06}). Although the [OIII]
emission-line constraints are typically obtained in a narrower slit
than the [OIV], since the majority of the [OIII] and [OIV] emission is
likely to be produced close to the AGN (i.e.,\ the ionising source),
aperture effects are probably not significant.

In Fig. \ref{fig_extinc}b we relate the two different measures of
dust-extinction within the host galaxies obtained using the Balmer
decrement and our [OIV]--[OIII] flux ratio estimate. Similar levels of
optical extinction are derived for the optically identified AGNs on
the basis of both the Balmer decrement and the [OIV]--[OIII] flux
ratios, implying moderate levels of dust obscuration ($A_V \sim 0$--3
mags). However, by contrast, although optically unidentified AGNs have
similar Balmer decrements to the optically identified AGNs, many have
significantly higher [OIV]--[OIII] flux ratios. This suggests that the
optically unidentified AGNs are so heavily extincted at optical
wavelengths that the Balmer decrement no longer provides a reliable
estimate of the amount of obscuration, suggesting $A_V>3$
mags. Indeed, on the basis of the [OIV]--[OIII] flux ratio for
NGC~4945 we estimate that the AGN is obscured behind a screen of $A_V
\sim 9$ magnitudes, which is in good agreement with the detailed
constraints of Moorwood et al. (1996). Additionally, \citet{dma99}
find using near-IR spectroscopy $A_V \sim 7$ magnitudes for NGC~5128,
which is again consistent with our extinction estimation derived from
the [OIV]--[OIII] flux ratio ($A_V \sim 5$ magnitudes). We predict
that if the [OIII] emission was adjusted for an additional absorption
within the narrow line region (NLR) as found by our diagnostic, these
heavily extinguished optically unidentified AGNs would be moved into
the Seyfert region of an optical BPT diagram. For example, assuming
there is no further reddening to the H$\beta$ emission, an additional
NLR extinction of $A_V \sim 3$ is consistent with an increase by a
factor of 6 in [OIII] flux (i.e., the correction required for NGC~5128
to be optically classified as a Seyfert galaxy; see
Fig.~\ref{fig_bpt}).

In Fig.~\ref{fig_oiii_nev}a, we show that three of the optically
unidentified AGNs (NGC~0613, NGC~3627 and NGC~5195) are within the
intrinsic scatter of the [OIII]-[NeV] relation (shaded region), and
consequently do not appear to be deficient in [OIII] flux. Indeed, in
Fig. \ref{fig_extinc}b we find that these particular objects do not
appear to be heavily extinguished at optical wavelengths ($A_V < 3$
mags). It is therefore likely that these are not classified as AGNs in
optical surveys due to enhanced H$\beta$ flux from young stars, which
lowers the [OIII]/H$\beta$ flux ratio in a BPT diagram (see
Fig.~\ref{fig_bpt}). This is in agreement with the AGN--star-formation
ratios estimated for these objects with {\it Spitzer}-IRS; see
Fig.~6b. However, it is clear from Fig.~\ref{fig_oiii_nev}a,b that the
majority of the optically unidentified AGNs are highly extinguished in
[OIII] $\lambda 5007$\AA \ flux.

\begin{figure*}
\begin{center}
\vspace{-1.5cm}
\includegraphics[width=0.9\textwidth]{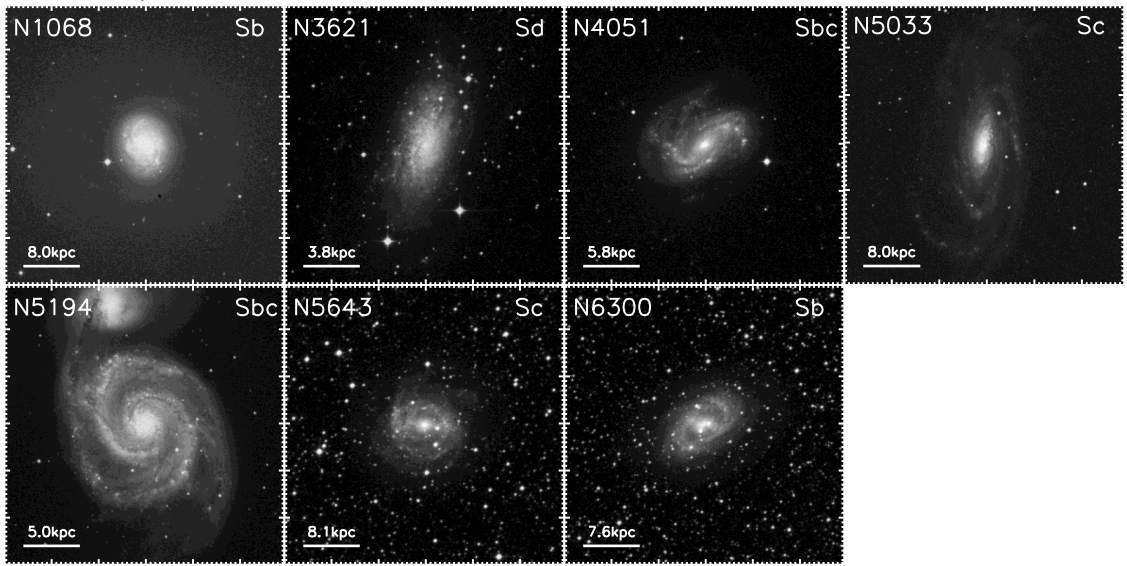}
\vspace{-1.8cm}
\caption{$10\times10$ arcmin ESO Digital Sky Survey images (combined
$B$ and $R$ band) of the optically unidentified and optically
identified AGNs. Optically unidentified AGNs are generally found to be
hosted in either highly inclined galaxies (E121-G006, NGC~660,
NGC~1448, NGC~3628 and NGC~4945) or galaxies with obscuring dust lanes
(NGC~1792, NGC~3628, NGC~4945 and NGC~5128). This suggests that the
lack of optical AGN signatures in these galaxies could be due to
extinction through the host galaxy, indicating that obscuration of the
nucleus is not necessarily due to an obscuring torus. Three of the
optically unidentified AGNs (NGC~613, NGC~3627 and NGC~5195) appear to
be hosted in relatively face-on galaxies; we suggest that the AGN
optical signatures are diluted in these galaxies due to strong
star-formation activity (see \S3.2.2 \& \S3.2.3).}
\label{fig_agn_images}
\end{center}
\end{figure*}

\subsection{First-order constraints on the source of the optical extinction}

The obscuration towards some of the AGNs implied by
Fig.~\ref{fig_extinc}a,b may be due to either a dusty torus (as
predicted by the unified model for AGNs; e.g.,\ \citealt{antonucci93})
or extinction through the host galaxy (e.g., \citealt{malkan98}; Matt
2000). Here we can test the latter hypothesis by looking for evidence
of host-galaxy extinction (i.e.,\ a highly inclined galaxy or
obscuring dust lanes) using available optical imaging; we defer
exploration of the former hypothesis to Goulding et~al. (in prep.)
where we search for evidence of nuclear obscuration in our sample
using sensitive X-ray observations. In Fig. \ref{fig_agn_images} we
compare optical images of the optically identified and unidentified
AGNs. The optically identified AGNs in our sample are hosted by
relatively face-on galaxies while, by contrast, many of the optically
unidentified AGNs are hosted in either highly inclined galaxies or
have dust lanes that obscure the central regions. Indeed, the four
optically unidentified AGNs with the highest [OIV]--[OIII] flux
ratios, which imply $A_V>3$~mags, are all hosted by galaxies that are
either highly inclined or have obscuring dust lanes; ESO121-G006 does
not have good-quality optical spectroscopy, however, since it resides
in a highly inclined host galaxy, it is likely to be an optically
unidentified AGN with a high [OIV]--[OIII] flux ratio. Furthermore,
three of the optically unidentified AGNs are found to be hosted in
relatively face-on galaxies (NGC~0613, NGC~3627 and NGC~5195), similar
to that of the optically identified AGNs, suggesting there is little
extinction from the host galaxy towards these AGNs. This is consistent
with our findings in the previous section that these galaxies do not
appear to have heavily extinguished [OIII] emission (see
Fig.~\ref{fig_oiii_nev}a).

Our findings extend the analyses of Malkan et~al. (1998), who found
that dust lanes can dictate the observed optical AGN type, by now
showing that the host galaxy can have a large effect on even the {\it
identification} of AGN activity. We find that four ($\approx
40^{+31}_{-19}$ percent) of the ten optically unidentified AGNs reside
in highly inclined galaxies, compared to $\approx$~4 percent (Seyfert
1) and $\approx$~7 percent (Seyfert 2) of the optically identified
AGNs in Malkan et~al. (1998), respectively. Four ($\approx
40^{+31}_{-19}$ percent) of the optically unidentified AGNs also
appear to have obscuring dust lanes, compared to $\approx$~10 and
$\approx$~20 percent of the Seyfert 1 and Seyfert 2 galaxies in Malkan
et~al. (1998), respectively.\footnote{We note that high-resolution
{\it HST} observations are required for the optically unidentified
AGNs to provide a consistent comparison to Malkan et~al. (1998).}
Whilst these findings are empirical and may be subject to small number
statistics, there appears to be strong evidence to suggest that the
non-identification of optical AGN signatures in the majority of these
galaxies is due to extinction through the host galaxy, indicating that
obscuration of the nucleus is not necessarily due to an obscuring
torus. We also note that similar conclusions have been proposed for
the non-identification of AGN signatures in $z\approx$~0.5--1 X-ray
identified AGNs (e.g.,\ \citealt{rigby06}).

\subsection{Can further AGNs be identified using other mid-IR emission-line diagnostics?}
\label{sec:prev_diags}

In our analyses thus far we have conservatively identified AGNs using
the \nev emission line. However, there may be further AGNs in our
sample that can be identified using different mid-IR emission
lines. Other studies have used the high-excitation emission line [NeV]
$\lambda 24.32 \um$, as well as intermediate-excitation emission lines
such as [OIV] $\lambda 25.89 \um$ (54.9 eV) and [NeIII] $\lambda 15.56
\um$ (41.0 eV). To explore whether we can identify further AGNs in our
sample, we investigate these other potential AGN indicators here.

[NeV] $\lambda 24.32 \um$ has the same ionisation potential as \nev
but it has a higher critical electron density ($n_e \approx 5 \times
10^5$ versus $n_e \approx 5 \times 10^4 \pcm$; Sturm
et~al. 2002). [NeV] $\lambda 24.32 \um$ is detected in 11 of the 64
galaxies in our sample, all of which are also detected at [NeV]
$\lambda 14.32 \um$. The smaller AGN identification fraction found
using [NeV] $\lambda 24.32 \um$ may be due to the lower relative
sensitivity of data in the LH module in some cases. We also note that
a large number of pixels around $\lambda \approx 24.3 \um$ were
damaged by solar flares early in the {\it Spitzer} mission, and are
flagged by the \spitzer \ pre-processing pipeline as either damaged or
unreliable (`hot' pixels). In our custom reduction pipeline we removed
these pixels from the analysis and fitting routines to reduce the
potential for spurious detections. For example, from analysis of the
BCD images for NGC~3938 before and after cleaning using a custom
version of {\small IRSCLEAN}, we find that the possible [NeV] $\lambda
24.32 \um$ emission reported by S08 falls below our detection
threshold of $3 \sigma$. Furthermore, an additional `hot' pixel at
$\lambda \sim 24.3594 \um$ (in the observed frame) is not removed by
default in {\small IRSCLEAN} due to the high incidence of adjacent
cleaned pixels. During the reduction process, care must be taken to
remove such pixels to ensure reliable detections. Additionally, we do
not find any further AGNs in our $D<15$ Mpc sample on the basis of the
[NeV] $\lambda 24.32 \um$ emission line and urge caution when
identifying AGNs at low redshifts solely on the detection of weak
[NeV] $\lambda 24.32 \um$ with {\it Spitzer}-IRS
observations. Certainly, at higher redshifts (i.e.,\ where the
observed [NeV] $\lambda 24.32 \um$ emission is not affected by the
spurious pixels at $\lambda \sim 24.35 \um$) detections of weak [NeV]
$\lambda 24.32 \um$ without complimentary \nev will be possible for
systems with high levels of relative dust extinction; however, this is
likely to be rare as $A_{14.32}/A_{24.32} \sim 1.03$ \citep{chiar06}.

\oiv is often attributed to AGN activity. However, since it is an
intermediate excitation emission line it can also be detected in
galaxies experiencing heightened starburst activity (D06). \oiv is
detected in 41 of the 64 galaxies in our sample, including all of the
17 AGNs identified using [NeV]. For the mid-IR identified AGNs, we
find that \oiv and \nev emission are well correlated, suggesting that
\oiv is a good proxy for the intrinsic AGN luminosity in
[NeV]--identified AGNs; see Fig. \ref{fig_oiv_nev}. The regressional
fit is characterised by the equation given below, with a spread in the
data of only 0.24 dex (see also Footnote~9). We highlight the caveat
that this relation may only hold for the emission-line and IR
luminosity range explored here. To rigorously assess the validity of
this relation for all galaxies a further study probing a large
luminosity range must be undertaken, which is beyond the scope of
these analyses presented here.

\begin{figure}
\begin{center}
\hspace{-1.0cm}
\includegraphics[width=1.1\linewidth]{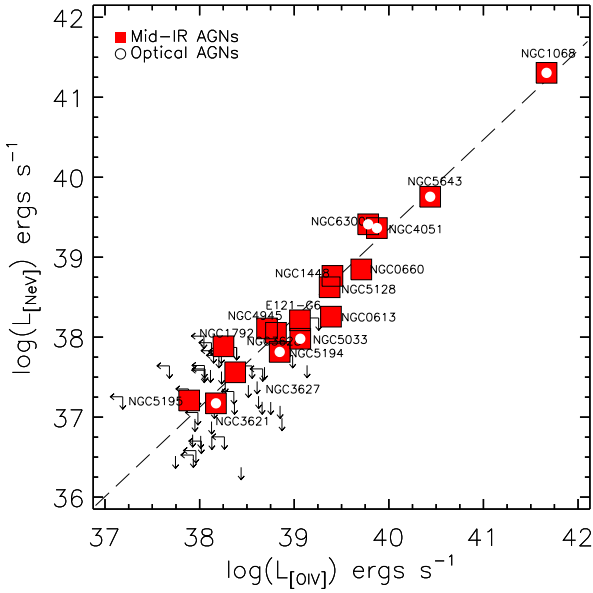}
\caption{\nev luminosity versus [OIV] $\lambda 25.89 \um$
luminosity. The dashed line was obtained using a robust bi-sector
linefit algorithm and indicates that for [NeV]--detected objects,
[OIV] $\lambda 25.89 \um$ provides a good proxy for the intrinsic
luminosity of the AGN (see Equation 2) for those galaxies with
$10^{39} < {\rm L}_{\rm [OIV]} < 10^{42} \ergps$. The symbols have
the same meaning as in Fig.~6.}
\label{fig_oiv_nev}
\end{center}
\end{figure}

\begin{figure}
\begin{center}
\hspace{-1.0cm}
\includegraphics[width=1.1\linewidth]{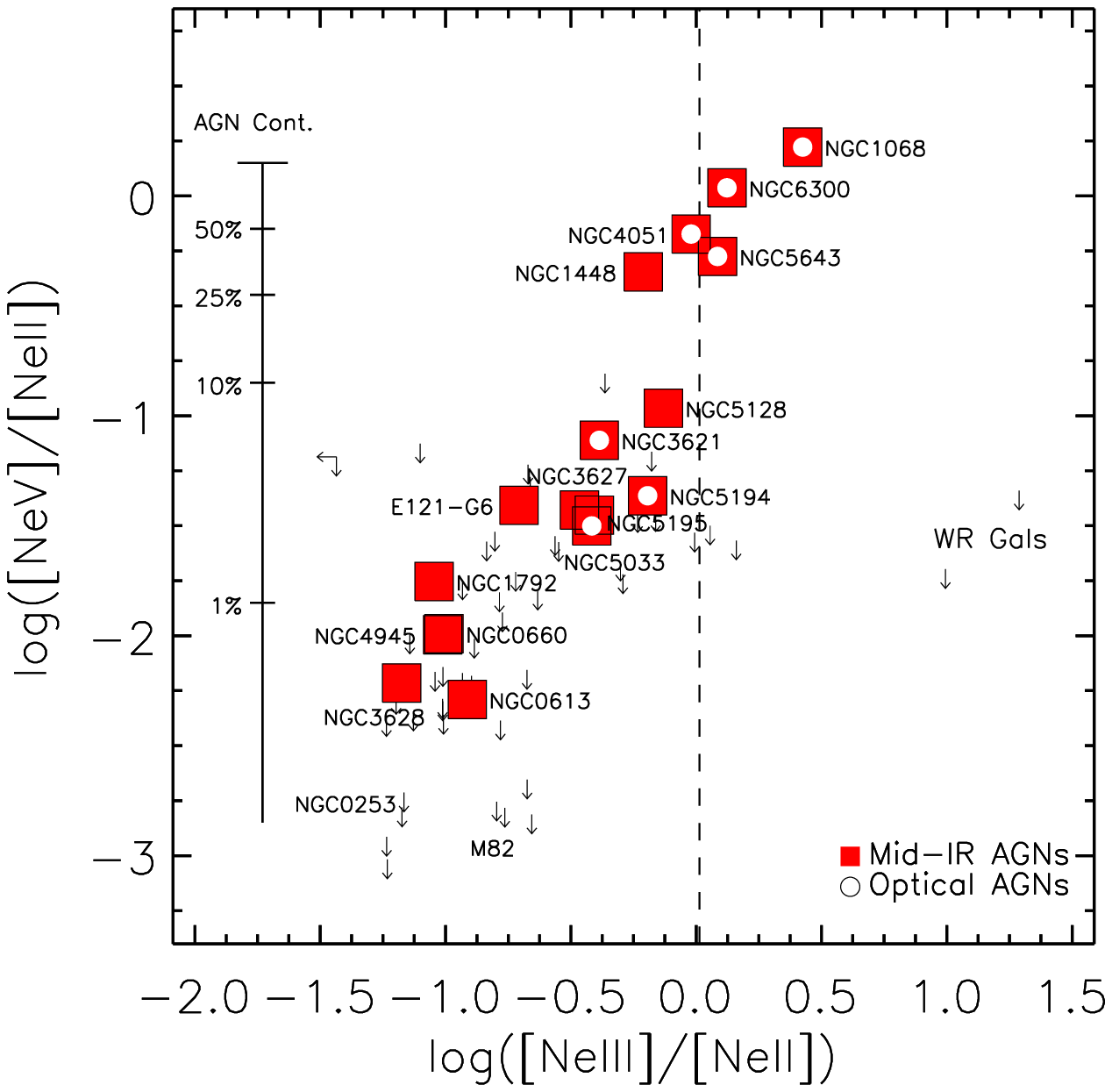}
\caption{Logarithm of the ratio of the fluxes of [NeIII] $\lambda
  15.51 \um$ and [NeII] $\lambda 12.82 \um$ emission lines are plotted
  against the logarithm of the ratio of the fluxes of \nev and [NeII]
  $\lambda 12.82 \um$ emission lines. The dashed line represents the
  demarcation for Seyfert 2 galaxies given in Melendez et~al.
  (2008). Those AGNs which dominate the luminosity output of the
  galaxy have preferentially higher [NeIII]/[NeII] ratios, and appears
  to be a good proxy for AGN activity in these cases. By contrast,
  lower luminosity AGNs exhibit similar ratios to that of star-forming
  galaxies. Wolf-Rayet galaxies introduce a further caveat to use of
  the [NeIII]/[NeII] ratio to indicate AGN activity,
  ([NeIII]/[NeII])$_{\rm AGN} \sim $ ([NeIII]/[NeII])$_{\rm WR}$. The
  symbols have the same meaning as in Fig.~6.}
\label{fig_neiii}
\end{center}
\end{figure}

\begin{equation}
{\rm log [NeV]} = (1.09 \pm 0.06) \ {\rm log [OIV]} - (4.34 \pm 2.53)
\end{equation}

Our finding of a tight relationship between \oiv and \nev suggests
that any AGN activity with bright \nev emission should also have
bright \oiv emission. Contrary to this, S08 identified two AGNs
(NGC~4321 and NGC~4536) using \nev that had undetected \oiv
emission. NGC~4536 is present in our sample but we do not identify
\nev using our conservative reduction and analysis techniques. Four of
the galaxies (IC2056; NGC~3175; NGC~3184; NGC~3368) in our sample with
\oiv detections lie above the regressional fit in
Fig. \ref{fig_oiv_nev} and more sensitive mid-IR spectroscopy may
identify AGN activity with \nev in these systems. A further 20
galaxies undetected in \nev lie below the regressional fit and,
therefore, unless AGNs can have weak \nev and bright \oiv emission,
these do not appear to host AGN activity. However, we note that the 13
galaxies in our sample that are undetected in both \oiv and \nev may
host AGN activity substantially below our detection
limits. Furthermore, for galaxies with L$_{\rm [OIV]} < 10^{39}
\ergps$, the source of the [OIV] emission becomes uncertain as a
non-negligible fraction may be produced by star-formation (e.g., the
starburst-AGN, NGC~4945). Therefore, the [OIV] emission may not be a
strong tracer of the intrinsic luminosity of the AGN (i.e., the [NeV]
luminosity) for galaxies with L$_{\rm [OIV]} \ll 10^{39}
\ergps$. Further probes of the AGN luminosity (e.g., hard X-ray
luminosities) are required to test this further and we defer this
analysis to Goulding et al. (in prep.).

[NeIII] $\lambda 15.56 \um$ is also an intermediate excitation
emission line and could be produced by AGN or star-formation
activity. Previous surveys (e.g., \citealt{farrah07};
\citealt{groves08}; \citealt{melendez08a}) have selected potential
AGNs on the basis of high [NeIII]--[NeII] flux ratios
([NeIII]/[NeII]$> 1.03$; \citealt{melendez08a}). In
Fig.~\ref{fig_neiii} we show that while this diagnostic identifies
many of the AGN-dominated systems in our sample, it does not find all
of the AGNs. The production of [NeIII] appears to be complex and may
have strong dependences on star-formation, gas density, temperature,
metallicity and AGN activity (Brandl et~al. 2006); for example, there
is a similar range of [NeIII]--[NeII] ratios for both the optically
unidentified AGNs and the galaxies without clear signatures of AGN
activity. Furthermore, the two WR galaxies in our sample also have
high [NeIII]--[NeII] flux ratios, suggesting that further criteria are
required to unambiguously identify AGN activity. However, three of our
galaxies not identified as AGNs (NGC~4490; NGC~4736; NGC~6744) have
high [NeIII]--[NeII] flux ratios and may host AGN activity below our
\nev sensitivity limit.

\section[]{Conclusions}

We have presented the initial results of a sensitive volume-limited
\spitzer-IRS spectral survey of all ($\approx 94$ percent)
bolometrically luminous ($\Lir \goa 3 \times 10^9 \Lsun$) galaxies to
$D < 15$ Mpc. We place direct constraints on the ubiquity of AGN
activity in the local universe. Our main findings are the following:

\begin{enumerate}
\renewcommand{\theenumi}{(\arabic{enumi})}

\item By conservatively assuming that the detection of the
high-excitation \nev emission line indicates AGN activity, we
identified AGNs in 17 of the 64 galaxies in our sample. This
corresponds to an AGN fraction of $\approx 27^{+8}_{-6}$ percent, a
factor $\goa 2$ greater than found using optical spectroscopy alone;
further AGNs are likely to be identified below our \nev sensitivity
limit. The large AGN fraction indicates a tighter connection between
AGN activity and IR luminosity for galaxies in the local Universe than
previously found, potentially indicating a close association between
AGN activity and star formation.

\item We explored whether the absence of optical AGN signatures in the
optically unidentified AGNs is due to either an intrinsically
low-luminosity AGN or dilution from star formation. We found that the
optically unidentified AGNs are typically characterised as
star-formation dominated galaxies hosting modest-luminosity AGNs
($\Lnev \approx 10^{37}$--$10^{39} \ergps$). However, since about half
of the optically identified AGNs also have these properties, it seems
unlikely that the dominant reason for the absence of optical AGN
signatures is due to either an intrinsically low-luminosity AGN or
overwhelming star-formation activity. Indeed, we find that only three
of the optically unidentified AGNs may have enhanced H$\beta$ emission
from young stars which could dilute the optical AGN signatures. We
also showed that the absence of optical AGN signatures is not due to
low-sensitivity optical data. See \S3.2.1, \S3.2.2, \& \S3.2.3.

\item We explored whether the absence of optical AGN signatures in the
optically unidentified AGNs is due to optical extinction. We found
that the optically unidentified AGNs typically have larger [OIV]
$\lambda 25.89 \um$--[OIII] $\lambda 5007$\AA \ flux ratios than the
optically identified AGNs, indeed suggesting that their emission is
typically heavily extinguished at optical wavelengths ($A_{\rm
V}\approx$~3--9 mags). Examination of optical images revealed that
seven of the optically unidentified AGNs are hosted in highly inclined
galaxies or galaxies with dust lanes, indicating that obscuration of
the AGN is not necessarily due to an obscuring torus. See \S3.2.3 and
\S3.3.

\end{enumerate}

We therefore conclude that optical spectroscopic surveys miss
approximately half of the AGN population simply due to extinction
through the host galaxy. Sensitive X-ray observations are required to
determine what fraction of the optically unidentified AGNs are heavily
obscured due to nuclear obscuration intrinsic to the AGN.

\section*{Acknowledgments}

We thank the referee for a considered and thorough report which has
significantly improved the paper. We would also like to acknowledge
useful conversations with A. Edge, J. Geach, B. Lehmer,
M. Mel{\'e}ndez, J. Mullaney, and I. Smail. ADG is supported by the
award of a Science \& Technologies Facilities Council (STFC)
studentship and DMA is funded by a Royal Society fellowship and the
Leverhulme Trust. This research has made use of the NASA/IPAC
Extragalactic Database (NED) and the NASA {\it Spitzer} Space
Telescope which are operated by the Jet Propulsion Laboratory,
California Institute of Technology, under contract with the National
Aeronautics and Space Administration. We would like to acknowledge the
use of the Sloan Digital Sky Survey (SDSS) data archive, funding
provided from the Alfred P. Sloan Foundation.

\bibliography{bibtex1}

\bsp 

\label{lastpage}

\begin{figure*}
\begin{center}
\includegraphics[width=0.8\textwidth,height=22.0cm]{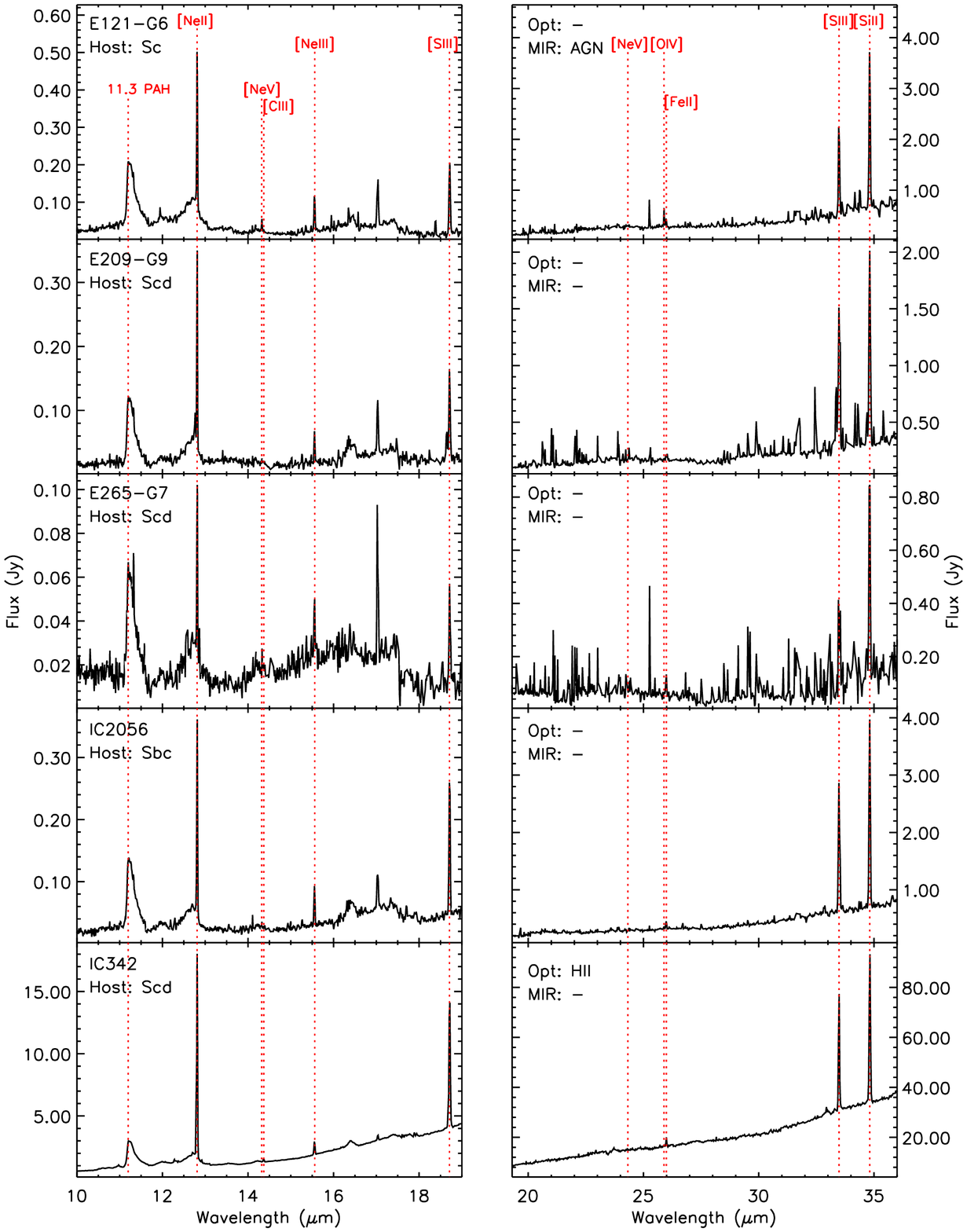}
\caption{\spitzer-IRS high-resolution spectra for the SH and LH
modules (left and right panels respectively) of all galaxies with {\it
Spitzer}-IRS observations in our $D<15$ Mpc IR-bright sample. The most
prominent spectral features are labeled and indicated with dotted
lines. The galaxy name, host-galaxy classification, optical spectral
classification, and mid-IR classification are indicated.}
\label{fig_spectra01}
\end{center}
\end{figure*}

\begin{figure*}
\begin{center}
\hspace{2.0cm}
\includegraphics[width=0.8\textwidth,height=22.0cm]{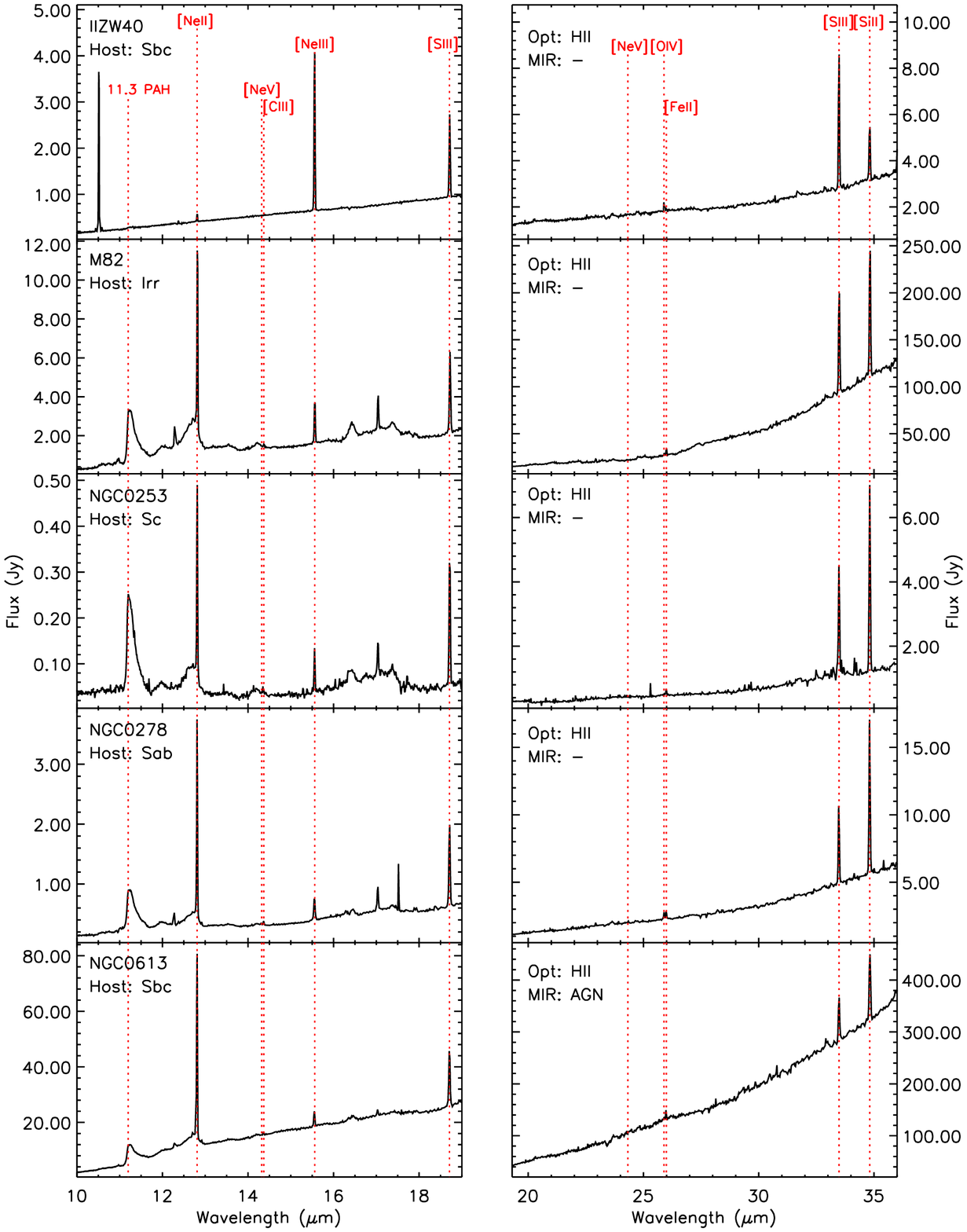}
\contcaption{}
\label{fig_spectra02}
\end{center}
\end{figure*}

\begin{figure*}
\begin{center}
\hspace{2.0cm}
\includegraphics[width=0.8\textwidth,height=22.0cm]{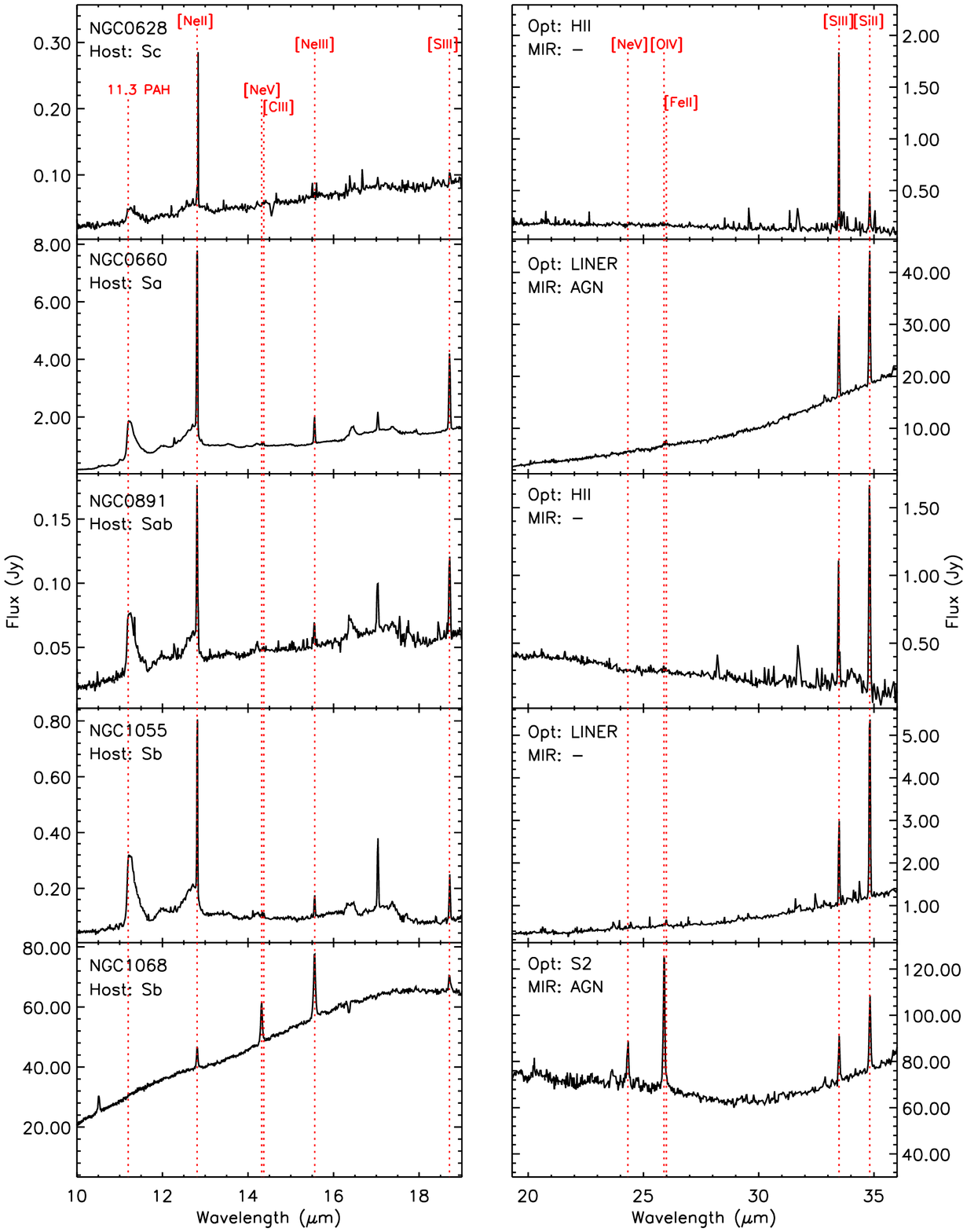}
\contcaption{}
\label{fig_spectra03}
\end{center}
\end{figure*}

\begin{figure*}
\begin{center}
\hspace{2.0cm}
\includegraphics[width=0.8\textwidth,height=22.0cm]{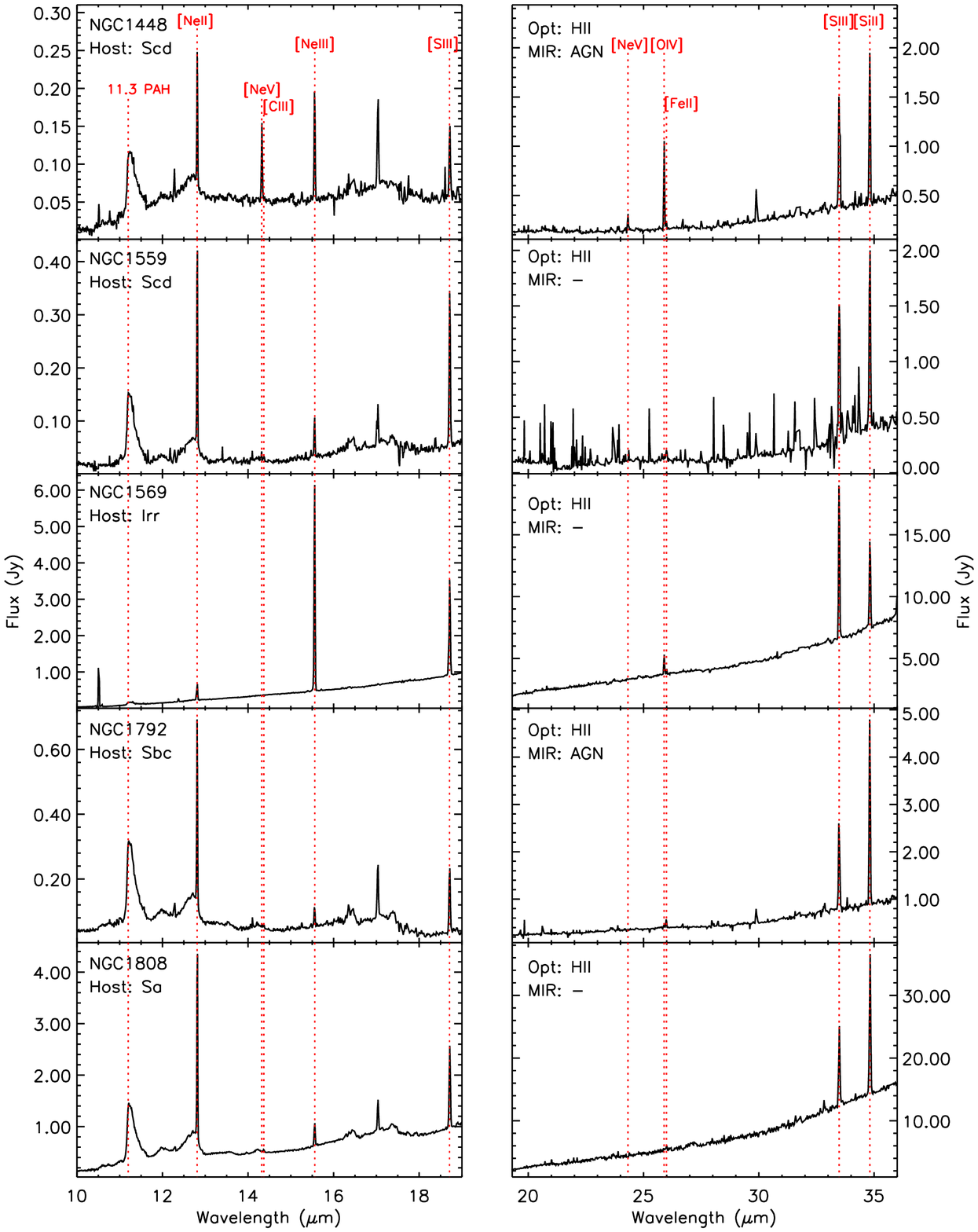}
\contcaption{}
\label{fig_spectra04}
\end{center}
\end{figure*}

\begin{figure*}
\begin{center}
\hspace{2.0cm}
\includegraphics[width=0.8\textwidth,height=22.0cm]{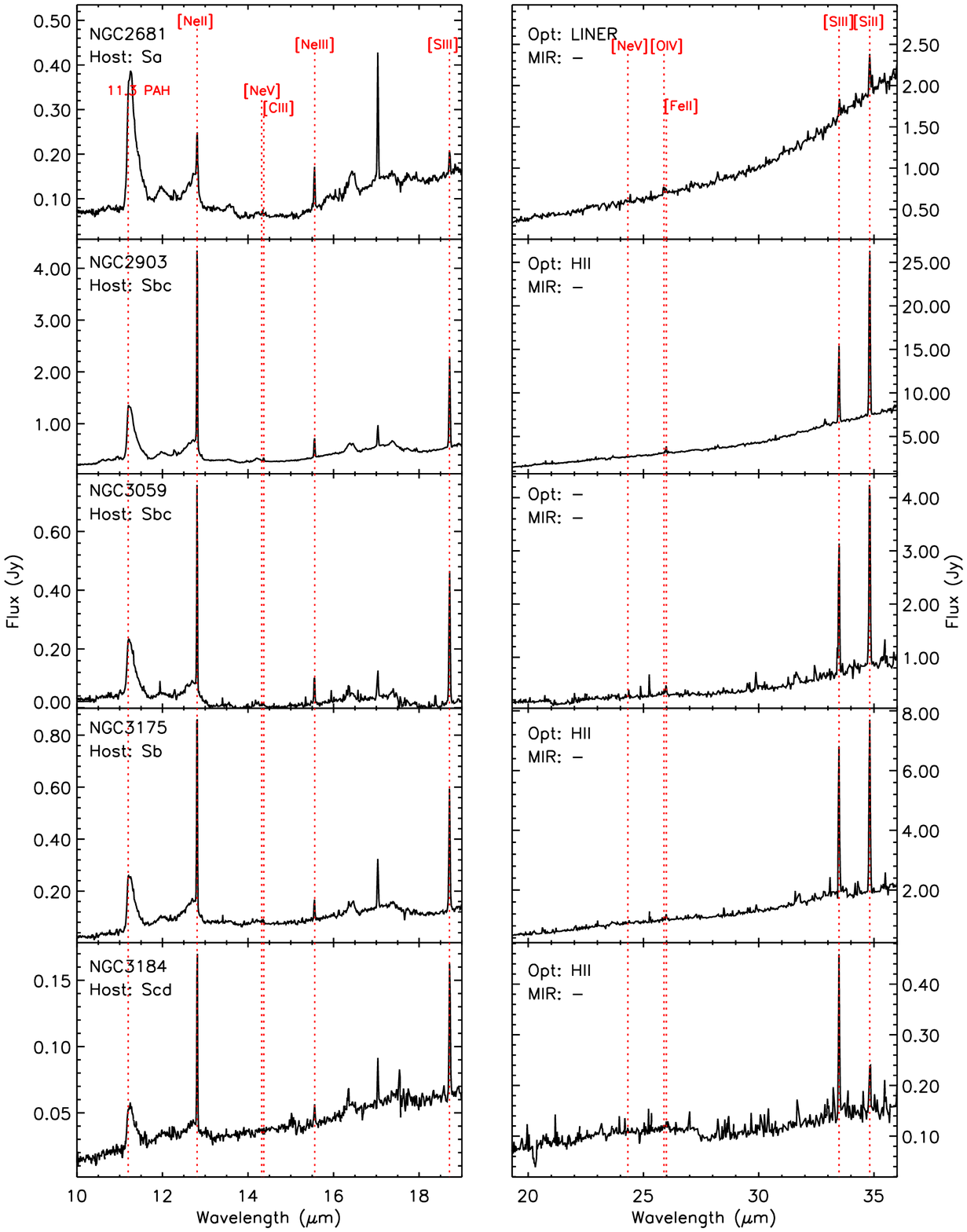}
\contcaption{}
\label{fig_spectra05}
\end{center}
\end{figure*}

\begin{figure*}
\begin{center}
\hspace{2.0cm}
\includegraphics[width=0.8\textwidth,height=22.0cm]{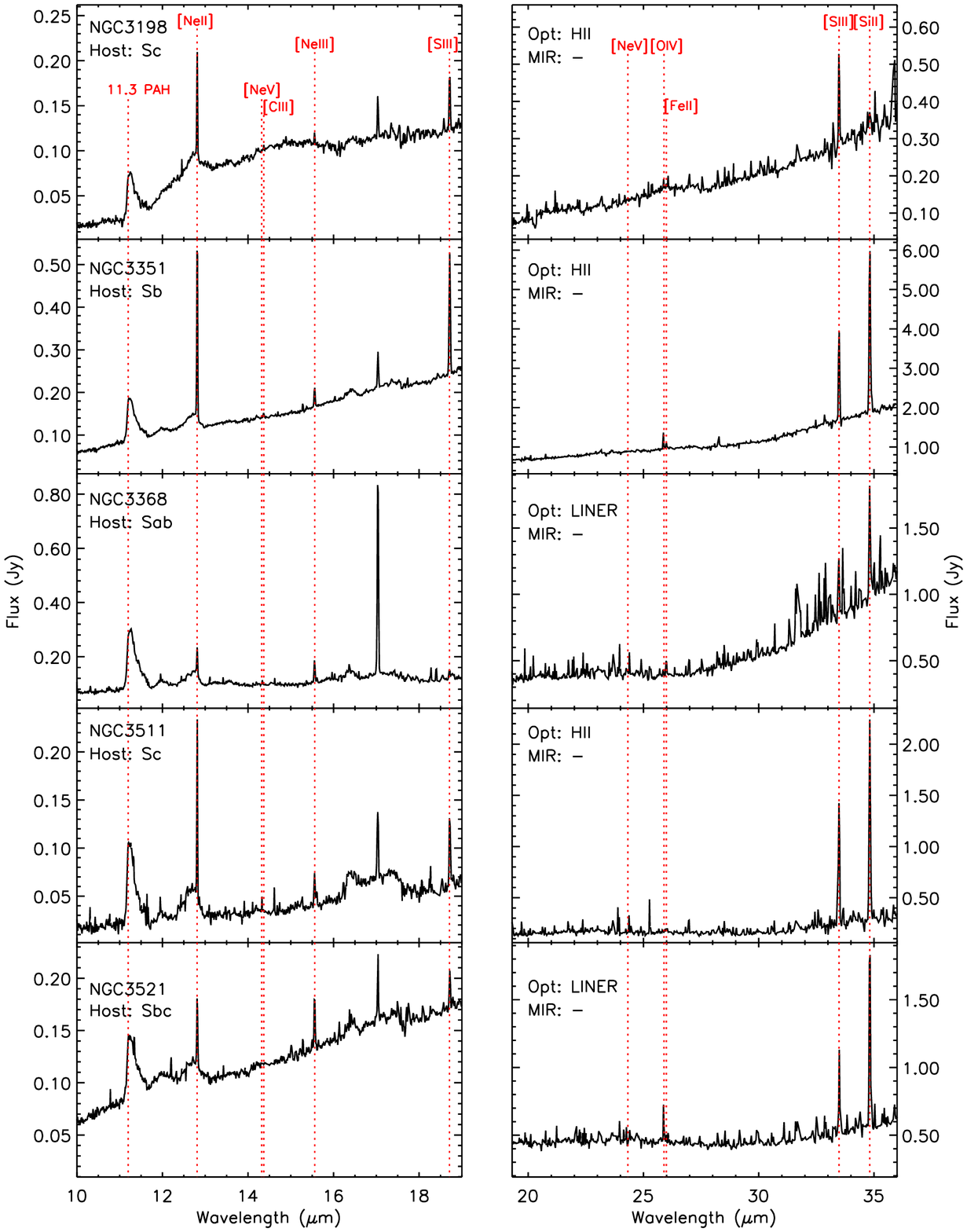}
\contcaption{}
\label{fig_spectra06}
\end{center}
\end{figure*}

\begin{figure*}
\begin{center}
\hspace{2.0cm}
\includegraphics[width=0.8\textwidth,height=22.0cm]{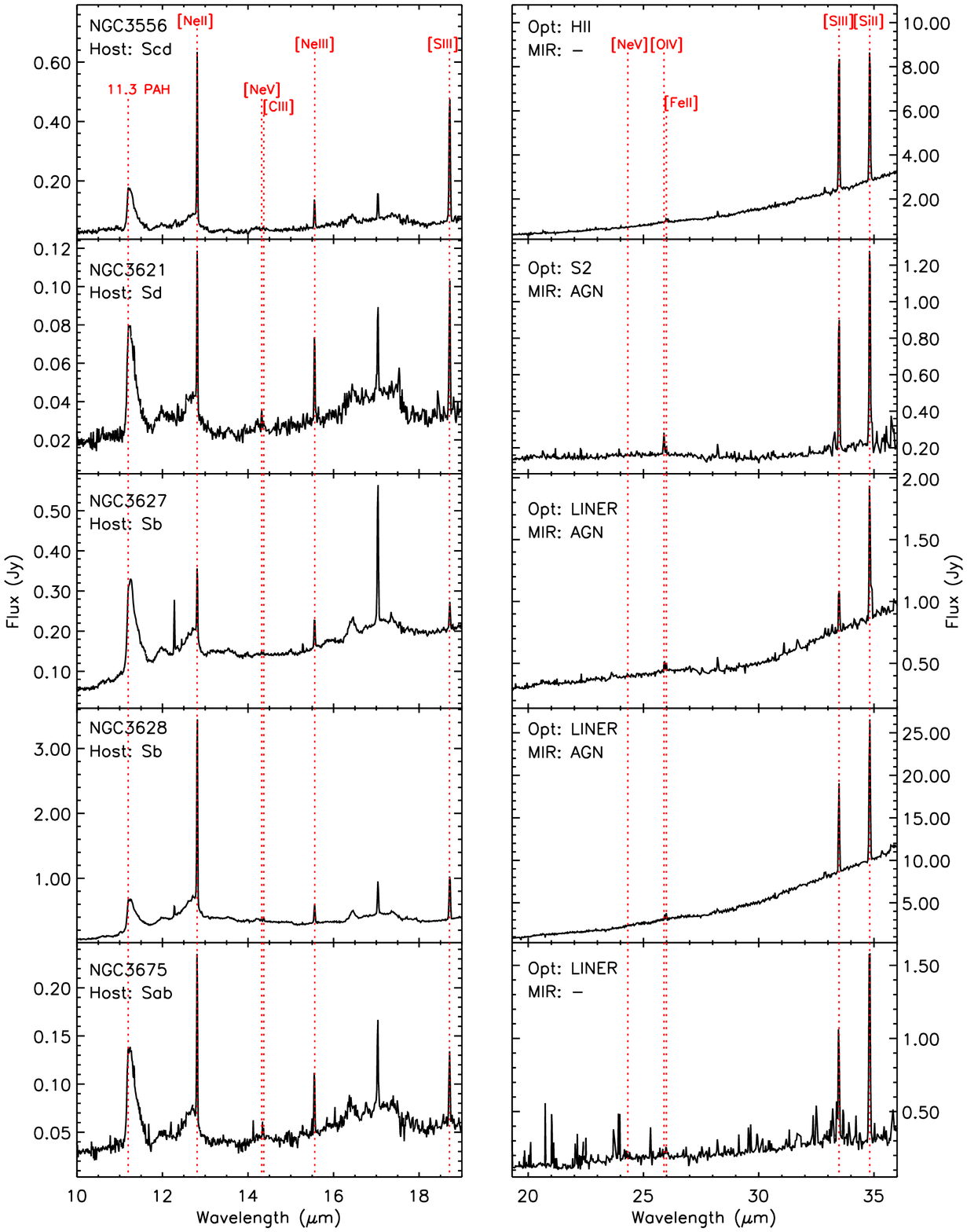}
\contcaption{}
\label{fig_spectra07}
\end{center}
\end{figure*}

\begin{figure*}
\begin{center}
\hspace{2.0cm}
\includegraphics[width=0.8\textwidth,height=22.0cm]{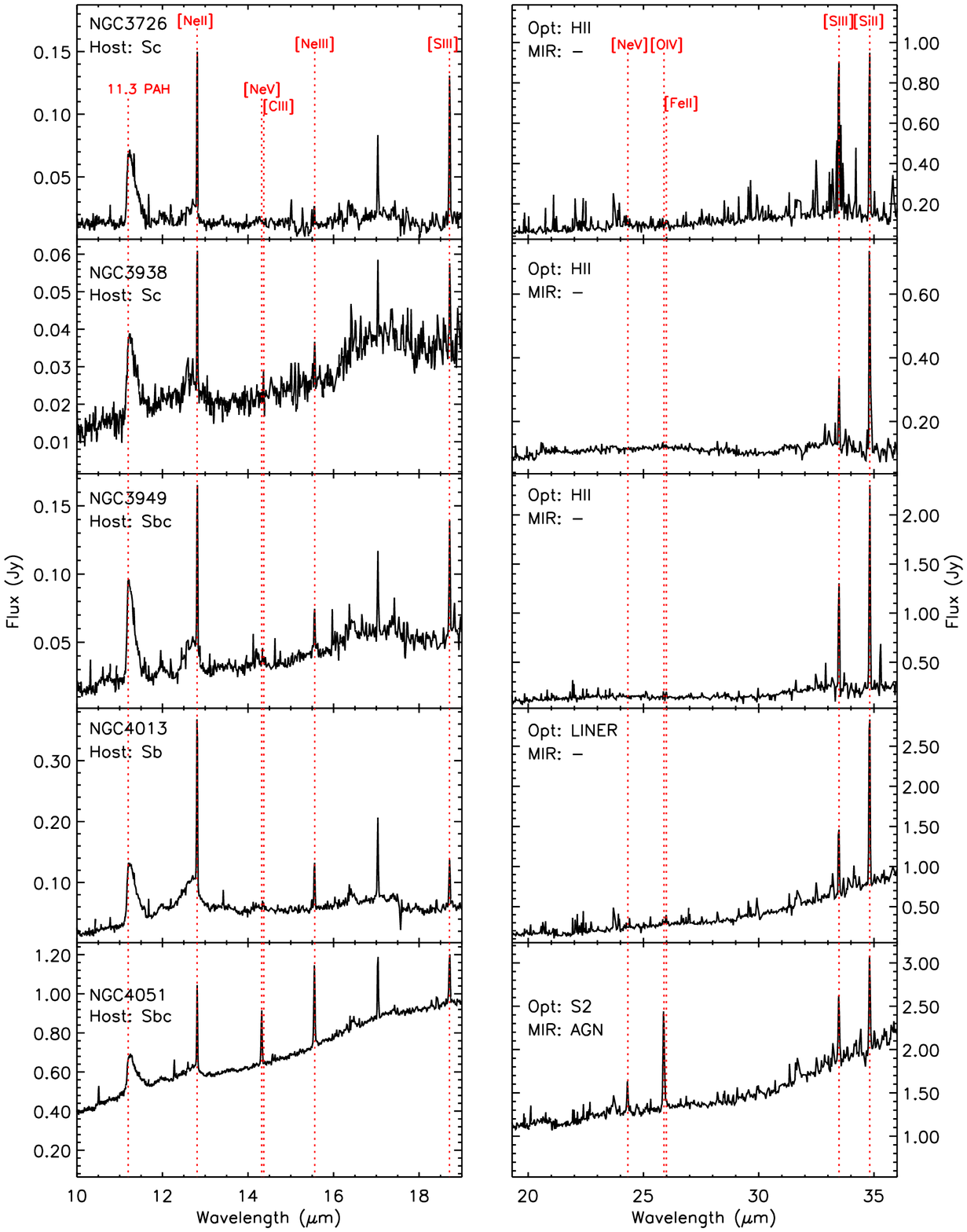}
\contcaption{}
\label{fig_spectra08}
\end{center}
\end{figure*}

\begin{figure*}
\begin{center}
\hspace{2.0cm}
\includegraphics[width=0.8\textwidth,height=22.0cm]{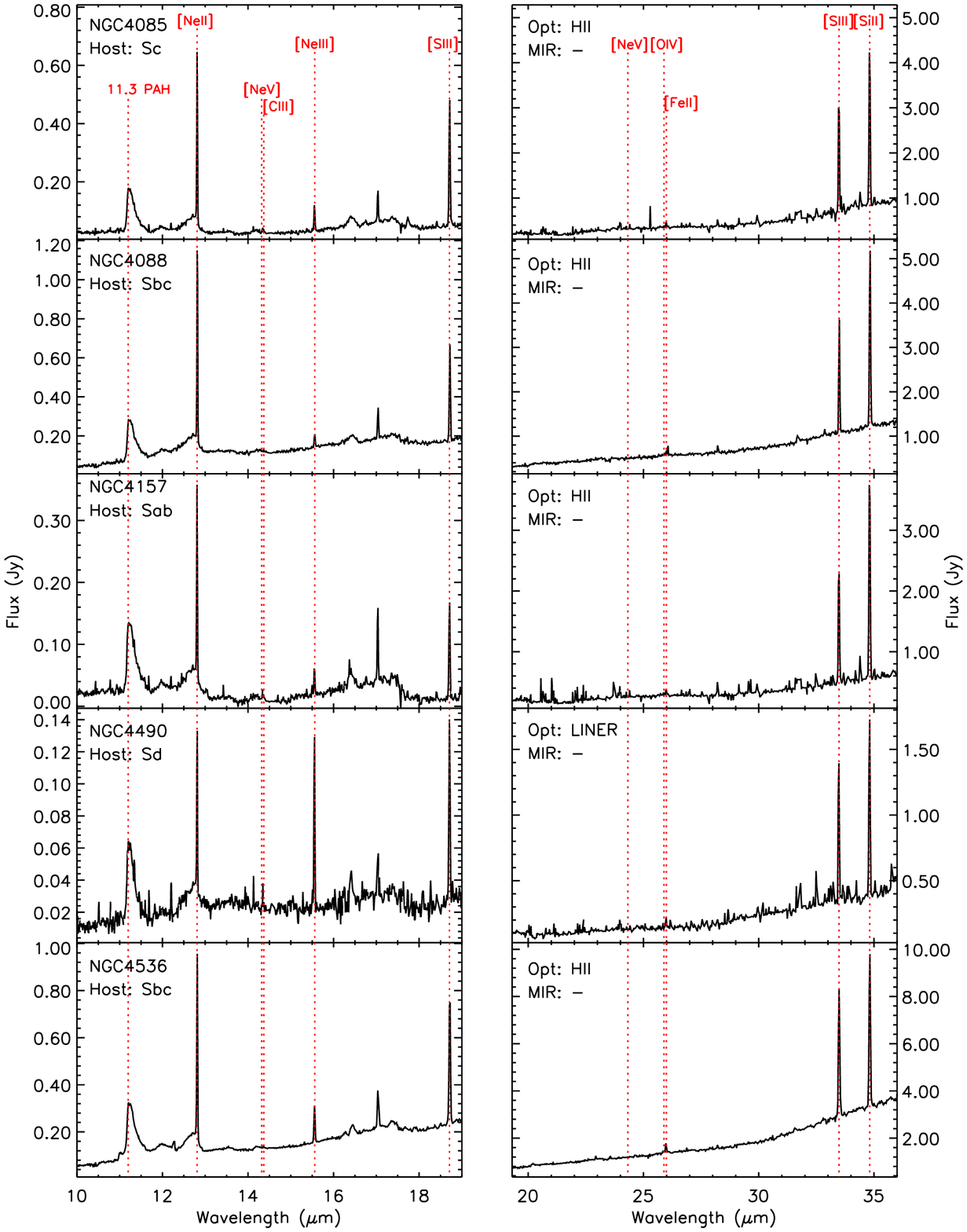}
\contcaption{}
\label{fig_spectra09}
\end{center}
\end{figure*}

\begin{figure*}
\begin{center}
\hspace{2.0cm}
\includegraphics[width=0.8\textwidth,height=22.0cm]{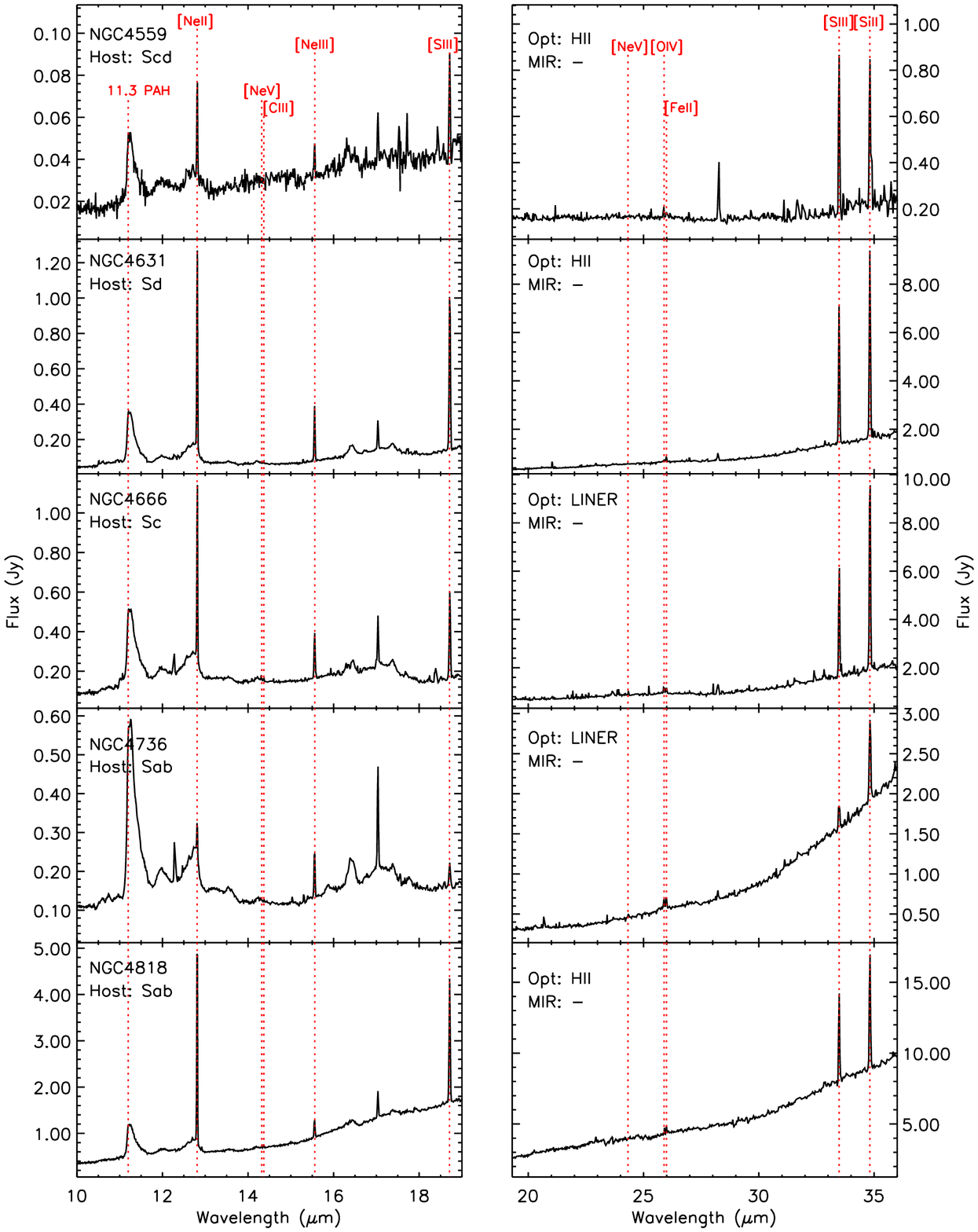}
\contcaption{}
\label{fig_spectra10}
\end{center}
\end{figure*}

\begin{figure*}
\begin{center}
\hspace{2.0cm}
\includegraphics[width=0.8\textwidth,height=22.0cm]{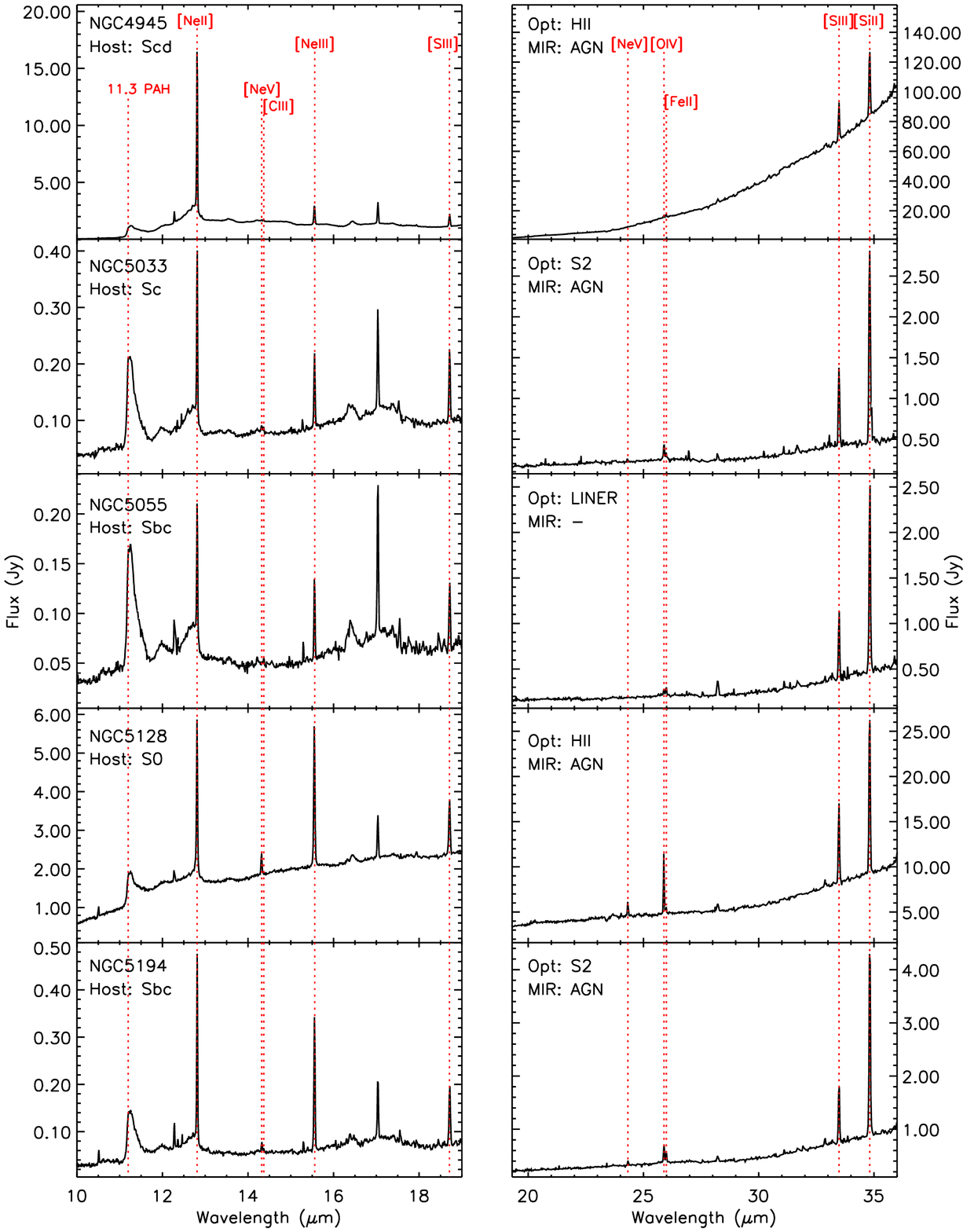}
\contcaption{}
\label{fig_spectra11}
\end{center}
\end{figure*}

\begin{figure*}
\begin{center}
\hspace{2.0cm}
\includegraphics[width=0.8\textwidth,height=22.0cm]{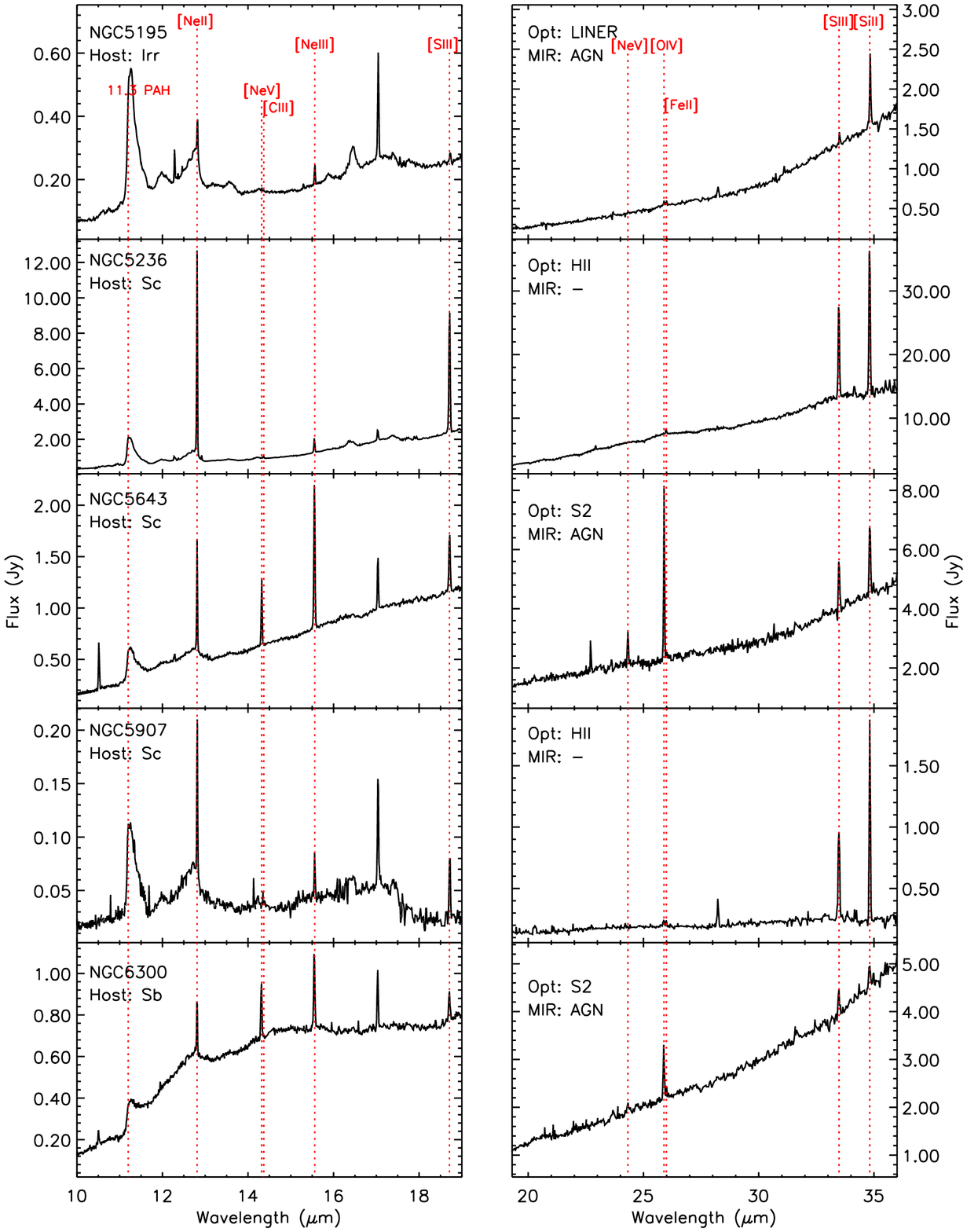}
\contcaption{}
\label{fig_spectra12}
\end{center}
\end{figure*}

\begin{figure*}
\begin{center}
\hspace{2.0cm}
\includegraphics[width=0.8\textwidth,height=22.0cm]{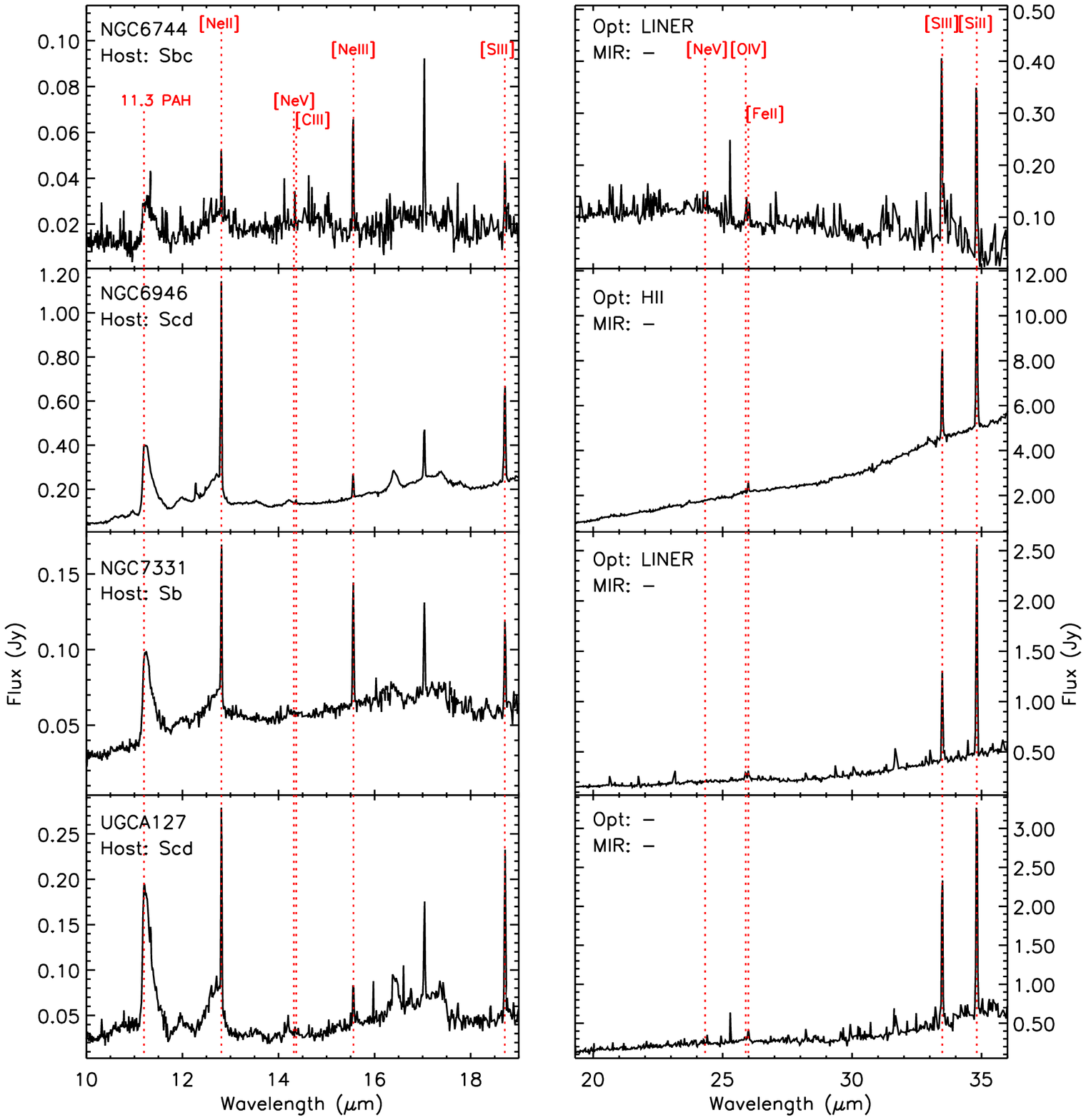}
\contcaption{}
\label{fig_spectra13}
\end{center}
\end{figure*}

\end{document}